\documentclass[twocolumn,pra,nolongbibliography,floatfix,aps,superscriptaddress,superscriptaddress]{revtex4-2}
\usepackage{amsthm,amsmath,amssymb,stmaryrd,amsfonts,graphicx,verbatim,xcolor,bm} 
\usepackage{ulem,enumerate} 
\usepackage[colorlinks=true,
            linkcolor=blue,
            citecolor=blue,
            urlcolor=blue]{hyperref}
\usepackage[utf8]{inputenc} 
\usepackage{siunitx,physics} 
\usepackage[english]{babel}
\usepackage{tabularx}
\usepackage{dsfont}

\usepackage{csquotes}
\MakeOuterQuote{"}

\newcommand{\ag}[1]{\textcolor{blue}{#1}}
\newcommand{\agx}[1]{\textcolor{gray}{\sout{#1}}}

\begin{document}

\preprint{APS/123-QED}

\title{  }

\author{Po-Jui Chen}
\affiliation{Department of Physics and Astronomy, Center for Materials Theory, Rutgers University, Piscataway, NJ 08854, USA}
\author{Kevin P. Lucht}
\affiliation{Department of Physics and Astronomy, Center for Materials Theory, Rutgers University, Piscataway, NJ 08854, USA}
\author{Andreas Gleis}
\affiliation{Department of Physics and Astronomy, Center for Materials Theory, Rutgers University, Piscataway, NJ 08854, USA}
\author{Andrew J. Millis}
\affiliation{Department of Physics, Columbia University, New York, NY 10027, USA}
\affiliation{Center for Computational Quantum Physics, Flatiron Institute, 162 5th Avenue, New York, NY 10010}
\author{J. H. Pixley}
\affiliation{Department of Physics and Astronomy, Center for Materials Theory, Rutgers University, Piscataway, NJ 08854, USA}
\affiliation{Center for Computational Quantum Physics, Flatiron Institute, 162 5th Avenue, New York, NY 10010}
\author{Daniele Guerci}
\affiliation{Department of Physics, William \& Mary, Williamsburg, VA-23187, USA}

\title{Spin Loop Currents and Bulk-Edge Structure in Correlated Topological States of WSe$_2$/MoTe$_2$ Bilayers}

\date{\today}

\begin{abstract}

Motivated by recent experiments, we theoretically investigate AB-stacked WSe$_2$/MoTe$_2$ bilayers. Our theory incorporates valence fluctuations , local and extended (longer ranged) Hubbard interactions, including non-local interactions between carriers on the MoTe$_2$ layer. 
Within the parton mean field framework, we show that the non-local  interaction between MoTe$_2$ carriers can drive a transition between a topological Kondo semimetal(TKSM) and a topological Kondo insulator(TKI) at the filling fraction $\nu =2$. We attribute this transition to an emergent spin loop current that arises from the interplay between Ising spin-orbit coupling and non-local electron correlations. In addition, we investigate how the emergent spin loop current affects the edge transport and show that the two-point conductance exhibits a quantized value at certain bias voltages.

\end{abstract}

\maketitle

\section{Introduction}\label{sec1}

Heavy-fermion  physics arises when localized magnetic moments hybridize with itinerant conduction electrons. The hybridization gives rise to a competition between Kondo screening and magnetic ordering  that can lead to a wide range of strongly correlated phases, including heavy Fermi liquids,  magnetic order, unconventional superconductivity, and non-Fermi-liquid behavior, with quantum critical points separating the different phases~\cite{doniachKondoLatticeWeak1977,stewartHeavyfermionSystems1984,mathurMagneticallyMediatedSuperconductivity1998,pfleidererSuperconductingPhasesElectron2009,steglichFoundationsHeavyfermionSuperconductivity2016,wirthExploringHeavyFermions2016,kimuraOpticalObservationNonFermiliquid2006}. 
An additional ingredient is band topology, which can endow Kondo hybridization with chirality and generate quasiparticle bands with nontrivial Berry curvature, thereby enabling topological phases~\cite{dzeroTopologicalKondoInsulators2010,dzeroTheoryTopologicalKondo2012}.

In transition metal dichalcogenide (TMD) heterobilayers, the moir\'{e} superlattice originates from the lattice mismatch of the two different materials, while the difference in chemical environments of the  constituent layers leads to different bandwidths and interaction strengths. In the MoTe$_2$/WSe$_2$ system considered here, the MoTe$_2$ layer has narrower bands and stronger correlations than the WSe$_2$ layer, and it is possible to access a parameter regime in which the MoTe$_2$ layer is in a Mott insulating phase with a local moment on each lattice site, while the WSe$_2$ layer is described by a weakly correlated highly conductive band, providing a highly tunable realization of heavy fermion physics.  A further point is that  for MoTe$_2$/WSe$_2$ bilayers in the AB-stacking configuration, the distinct spin and angular momentum of the highest-lying valence band states in the two layers generates a chiral inter-layer exchange, enabling a realization of a chiral Kondo lattice Hamiltonian in a controllable semiconductor setting~\cite{guerciChiralKondoLattice2023}. This system therefore provides a promising platform for addressing unresolved questions surrounding potential topological Kondo insulator candidates in  SmB$_6$ \cite{menthMagneticSemiconductingProperties1969,liTwodimensionalFermiSurfaces2014,phelanCorrelationBulkThermodynamic2014,chowdhuryMixedvalenceInsulatorsNeutral2018} and YbB$_{12}$ \cite{hagiwaraSurfaceKondoEffect2016,xiangHallAnomalyQuantum2022,guptaTopologicalSurfaceStates2023}. 

The experimental phase diagram of the MoTe$_2$/Wse$_2$ moir\'{e} heterobilayer is rich. At a carrier concentration of one hole per moir\'{e} unit cell, variation of the inter-layer potential difference ("displacement field") drives transitions from  an antiferromagnetic correlated insulator state in which the holes reside in the MoTe$_2$ layer and are Mott-insulating, to a  quantum anomalous Hall insulator identified through transport measurements~\cite{liQuantumAnomalousHall2021}. Complementary magnetic circular dichroism measurements further revealed  that the quantum anomalous Hall state is a valley-coherent state~\cite{taoValleyCoherentQuantumAnomalous2024} consistent with a canted anti-ferromagnet~\cite{devakulQuantumAnomalousHall2022}. Further increasing the displacement field leads  to a mixed valent metal with carriers in both layers. Starting from the antiferromagnetic correlated insulator and tuning the filling slightly above one hole per moir\'e unit cell, creates a ferromagnetic insulator~\cite{zhaoEmergenceFerromagnetismOnset2024}, which upon further doping  evolves to a nonmagnetic heavy Fermi liquid~\cite{zhaoGatetunableHeavyFermions2023}, in which  the quasiparticles acquire an enhanced effective mass compared to those in the weakly correlated regime at $\nu=2+\delta>2$, while the low-temperature resistivity exhibits a Fermi-liquid-like $T^2$  temperature dependence. 
The localized moments also contribute to the Fermi surface volume and lead to an enlarged Fermi surface \cite{oshikawaTopologicalApproachLuttingers2000}. At further doping to filling of 2 holes per moir\'e unit cell, a topological insulator was found in a recent experiment, and interpreted as a topological Kondo insulator (TKI)~\cite{han2025evidencetopologicalkondoinsulating}. 

These experimental findings inspired intensive theoretical efforts on theory of AB stacked TMD bilayer. For AB-stacking, a chiral Kondo lattice model with momentum-dependent interlayer hybridization has been derived~\cite{guerciChiralKondoLattice2023}, predicting a heavy Fermi liquid and a  topological Kondo semimetal (TKSM) at commensurate filling $\nu=2$~\cite{xieTopologicalKondoSemimetals2025a,xieKondoEffectIts2024} with non-trivial topological properties. The TKSM features a direct gap, but no indirect gap in the spectrum, resulting in compensated electron and hole Fermi surfaces. A full gap can open through heterostrain-induced lattice disorder~\cite{guerciTopologicalKondoSemimetal2024a}, where atomic-scale distortions are amplified by the moir\'e superlattice, resulting in a TKI state. 

\begin{figure}
\includegraphics[width=1\linewidth]{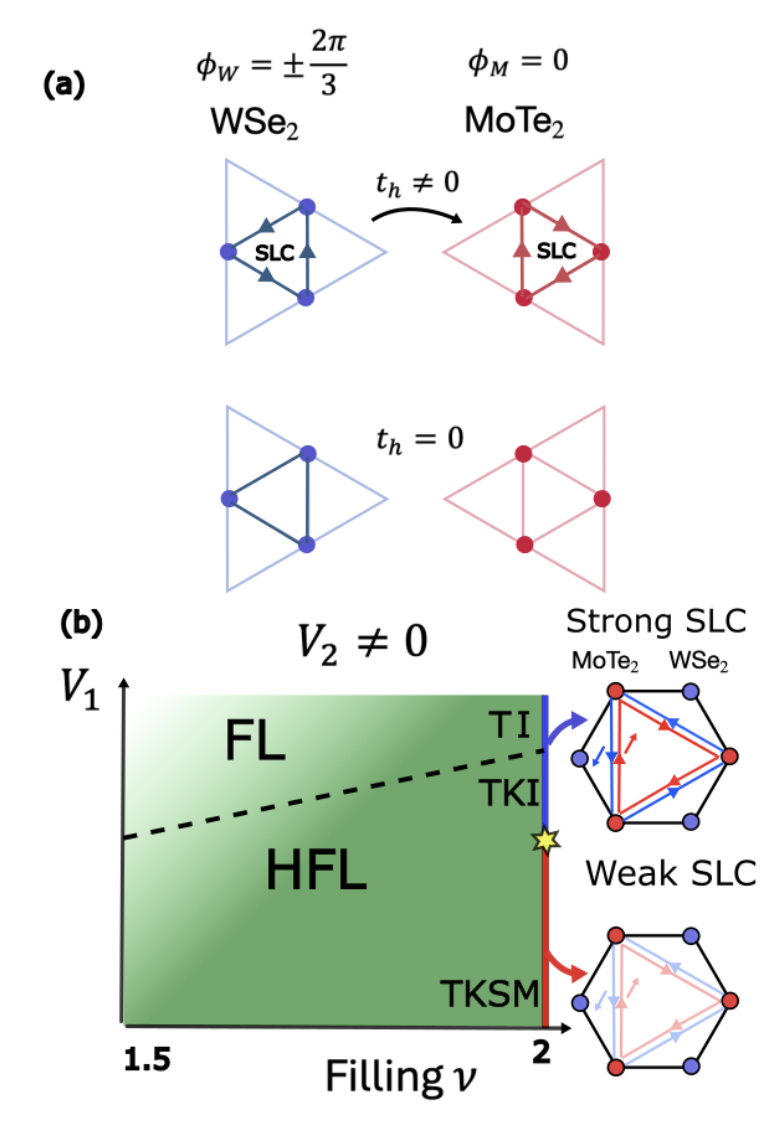}
    \caption{
 Schematic of (a) the proximity-induced spin-loop current (SLC) in TMD bilayer and (b) the phase diagram versus nearest-neighbor interaction $V_1$ and filling $\nu$. In (a), circulating arrows on the triangular lattices denote SLC mediated by interlayer hopping $t_h$. In (b), away from filling $\nu=2$, the ground state is a heavy Fermi liquid (HFL) at a small $V_1$, increasing $V_1$ gives rise to a Fermi liquid (FL). At $\nu=2$, increasing $V_1$ evolves the system from the topological Kondo semimetal (TKSM) toward the topological Kondo insulator (TKI). As $V_1$ increases further, there exsists a crossover from a TKI to a topological insulator (TI). The cartoons next to the phase diagram illustrate the weak and strong interaction-induced SLC regimes.  }    
   \label{schematic}
\end{figure}

Motivated by previous theoretical works, we complete the theoretical description of MoTe$_2$/WSe$_2$ heterobilayer by including the interaction between nearby WSe$_2$ and MoTe$_2$ carriers, $V_1$, and the interaction between two nearby MoTe$_2$ carriers, $V_2$~ \cite{sahaInterplayTopologyElectronelectron2025,sahaEmergentQuantumValley2026,morales-duranNonlocalInteractionsMoire2022}. 
In Ref.~\cite{gleis2026topologicalkondoinsulatorspin}, we have shown how the spin loop current (SLC) on Mo sites induced by $V_2$ appears within the dynamical mean field theory (DMFT). 
In related literature, loop current states have been proposed in chiral spin liquids~\cite{PhysRevB.39.11413}, cuprate superconductors~\cite{PhysRevB.55.14554} and topological Mott insulator~\cite{raghuTopologicalMottInsulators2008}.
Within the model introduced in Ref.~\cite{guerciChiralKondoLattice2023,guerciTopologicalKondoSemimetal2024a}, the emergence of SLC is mediated by the Ising spin-orbit coupling (SOC) of the heterostructure and enhanced by interactions. 
In the decoupling limit (see Fig.~\ref{schematic}~(a)), the WSe$_2$ and MoTe$_2$ layers act as independent triangular lattices of W and Mo, respectively. 
The WSe$_2$ layer hosts intralayer phase in the hopping.
When the inter-layer tunneling is present, this phase in the WSe$_2$ layer can be transferred to the MoTe$_2$ layer, which results in SLCs in that layer. Crucially, the nonlocal interaction $V_2$ enhances the intralayer hopping current in MoTe$_2$ and qualitatively changes the effective dispersion in the Mo layer, favoring a full gap in the spectrum and quantized spin Hall conductance in the TKI at $\nu=2$.

Here, we use parton mean-field theory to treat the non-local interactions, which allows us to examine the impact of the SLC as we vary total density, $\nu$.
At the mean-field level, we decouple the non-local Mo interaction which introduces the bond field, $Q$, modifying the intralayer hopping. 
Within a Ginzburg--Landau (GL) framework, we show explicitly that a complex-$Q$ solution is energetically favorable. The induced SLC not only develops at $\nu=2$ but also extends to finite doping ($1.5<\nu<2$), where  edge modes coexist with the bulk Fermi surface. Fig.~\ref{schematic} (b) shows a schematic phase diagram of the ground state as a function of filling $\nu$ and $V_1$, with constant $V_2$. Within our theoretical framework, we identify the existence of HFL, FL, TKSM and TKI in different parameter regions. The crossover from TKI to TI is not captured by our parton theory since our theory fail to capture the physics in a large $V_1$.

We investigate the edge-state properties, including the edge spectrum, group velocity, and transport characteristics for different filling factors. 
We show that the SLC alters the edge transport pattern, resulting in counter-propagating edge modes in the TKSM state. Moreover, the SLC-induced indirect gap enables the two-point conductance to become quantized at certain bias voltages, indicative of the transition from a topological semimetallic state to a topological insulating state.

The remainder of this paper is organized as follows. In Sec.~\ref{sec2}, we introduce the tight-binding model for the TMD heterobilayer and the interaction. Section~\ref{sec3} presents the slave-boson mean-field theory.
In addition, we also construct a finite temperature GL theory to rationalize our consideration of a complex bond field $Q$. Sec.~\ref{sec4} contains the self-consistently determined mean-field results for both incommensurate ($1.5 < \nu < 2.0$) and commensurate ($\nu=2$) fillings. We demonstrate the formation of the HFL or FL at incommensurate filling with nonlocal interactions, as well as the presence of a non-quantized Berry curvature. At commensurate filling, we show how the SLC produces a full gap, leading to a TKI defined by a spin Chern number. 
Sec.~\ref{sec5} focuses on edge modes. We examine the band structure under different boundary conditions and analyze the group velocities of edge states. Furthermore, we compute the local density of states (LDOS) and two-terminal conductance, both of which exhibit clear signatures of topological edge states. Finally, Sec.~\ref{sec6} summarizes the main results and discusses the connection between our theory and experiments.

\section{Tight-binding model}
\label{sec2}

\begin{figure}
\includegraphics[width=0.65\columnwidth]{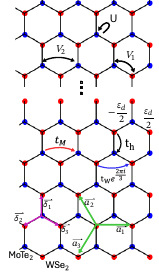}
    \caption{Lattice structure of AB-stacked WSe$_2$/MoTe$_2$ bilayer. The moir\'{e} superlattice is modeled as a honeycomb lattice composed of Mo on one sublattice site and W on the other. Lattice vectors are given by $\bm{a}_1$ and $\bm{a}_2$ with the set $\bm{a}_i$ for $i=1,2,3$ designating NNN latice vectors. NN lattice vectors are given as $\bm{\delta}_i$ for $i=1,2,3$. Hopping parameters between W and Mo are given as $t_\text{W}$ and $t_\text{M}$, respectively, $t_\text{h}$ is the interlayer tunneling strength between NN sites of Mo and W, and $\varepsilon_d/2$ with the appropriate sign for each layer corresponds to the displacement field they experience. Parameters $U,V_1$ and $V_2$ denote the interaction strength of nonlocal Hubbard interaction on Mo sites, interlayer Hubbard interaction and neighbor interaction between nearby Mo sites, respectively.}
   \label{TMD_lattice}
\end{figure}

The minimal model for the AB-stacked WSe$_2$/MoTe$_2$ bilayer is obtained by  Wannierizing  the two topmost moir\'{e} valence bands belonging to different layers~\cite{zhangSpintexturedChernBands2021a,devakulQuantumAnomalousHall2022}. 
For convenience, we adopt the shorthand notation W for WSe$_2$ and Mo for MoTe$_2$.
The two sets of Wannier centers for two interpenetrating triangular lattices, forming a honeycomb lattice with two inequivalent sublattices and with lattice vectors given by
$\bm{a}_{i+1} = a\big(\cos(\frac{2\pi}{3}i),\sin(\frac{2\pi}{3}i)\big)$ and $\bm{\delta}_{i+1} = \frac{a}{\sqrt{3}}\big(\sin(\frac{2\pi}{3}i),\cos(\frac{2\pi}{3}i)\big)$, with $i=0,1$ and $2$. (see Fig.~\ref{TMD_lattice}) 
The Hamiltonian~\cite{devakulQuantumAnomalousHall2022} reads

\begin{equation}
    \begin{split}
        \hat{H}= \hat{H}_c + \hat{H}_f+\hat{H}_{\text{h}}+\hat{H}_{\text{int}},
    \end{split}
    \label{eq:full_Ham}
\end{equation}
with $\hat{H}_c$ and $\hat{H}_f$ corresponding to the tight-binding Hamiltonians for the W and Mo layers, respectively, and $\hat{H}_{\text{h}}$ corresponding to the inter-layer hybridization Hamiltonian. The second quantized form of $\hat{H}_c,\hat{H}_f$ and $\hat{H}_\text{h}$ are given by:

\begin{widetext}
\begin{equation}\label{H_nonint}
    \begin{split}
        \hat{H}_c &= -t_\text{W}\sum_{\langle \bm r,\bm r'\rangle\in\text{W},\sigma}e^{i\frac{2\pi}{3}\nu_{\bm r,\bm r'}s_\sigma}c^{\dagger}_{\bm r,\sigma}c_{\bm r',\sigma}+\sum_{\bm r\in\text{W},\sigma}(\epsilon_d/2-\mu)c^{\dagger}_{\bm r,\sigma}c_{\bm r,\sigma},\\
        \hat{H}_f &= -t_\text{M}\sum_{\langle \bm r,\bm r'\rangle\in\text{Mo},\sigma}(f^{\dagger}_{r,\sigma}f_{\bm r',\sigma}+h.c)-
        \sum_{\bm r\in\text{Mo},\sigma}(\epsilon_d/2+\mu)f^{\dagger}_{\bm r,\sigma}f_{\bm r,\sigma},\\
        \hat{H}_{\text{h}} &= -t_\text{h}\sum_{ \bm r\in\text{W},\sigma,i}       (c^{\dagger}_{\bm r,\sigma}f_{\bm r+\delta_i,\sigma}+h.c.),
    \end{split}
\end{equation}
\end{widetext}
where $c_{\bm{r},\sigma}$ and $f_{\bm{r},\sigma}$ denote the fermion annihilation operators for W and Mo sites, respectively.  Parameters $t_\text{W},t_\text{M}$ and $t_{\text{h}}$ are the nearest-neighbor hopping strengths for W to W, Mo to Mo, and W to Mo, respectively. In our calculation, these values are taken to be $t_\text{W}=9$ meV, $t_\text{M}=4.5$ meV, and $t_\text{h}=2$meV~\cite{guerciChiralKondoLattice2023,guerciTopologicalKondoSemimetal2024a}. The parameter $\epsilon_d$ is the layer-biased potential, which can be tuned by applying a displacement field.

The inter-layer tunneling term displays nodes at the high-symmetry points $K$ and $K'$ of the mini-Brillouin zone and acquires a $p$-wave-like form around these nodes. In addition, the hopping process between two nearby W sites involves a phase factor $\frac{2\pi}{3}\nu_{\bm r\bm r'}s_{\sigma}$ which arises from the Ising spin orbit coupling on individual TMD layer~\cite{zhangSpintexturedChernBands2021a}. 
Here, $\nu_{\bm r,\bm r'}$ is antisymmetric under the exchange  of $\bm r$ and $\bm r'$. 
Its value is taken to be $\pm1$, depending on the path connecting $\bm r$ and $\bm r'$. 
The factor $s_\sigma =\pm 1$ imposes spin-dependent sign on the phase. 

The interaction in Eq.~\eqref{eq:full_Ham} is expressed as:
\begin{widetext}
\begin{equation}\label{H_int}
    \begin{split}
        \hat{H}_{\text{int}} &= U\sum_{\bm{r}\in\rm Mo}f^\dagger_{\bm{r},\uparrow}f_{\bm{r},\uparrow}
        f^\dagger_{\bm{r},\downarrow}f_{\bm{r},\downarrow} +
        V_1\sum_{\bm r, i,\sigma'\sigma}
        f^\dagger_{\bm{r}+\bm \delta_i,\sigma}f_{\bm{r}+\bm \delta_i,\sigma}
        c^\dagger_{\bm{r'},\sigma'}c_{\bm{r'},\sigma'} +V_2\sum_{\langle\bm{r,r'}\rangle\in\rm Mo,\sigma'\sigma}
        f^\dagger_{\bm{r},\sigma}f_{\bm{r},\sigma}
        f^\dagger_{\bm{r'},\sigma'}f_{\bm{r'},\sigma'}.
    \end{split}
\end{equation}
\end{widetext}

The three terms of Eq.~\eqref{H_int} consists of: on-site interaction on  Mo sites $U$, nearest-neighbor interaction between W and Mo sites $V_1$, and next-nearest-neighbor interaction between nearby Mo sites $V_2$. Based on the Wannierization, the magnitude of $U, V_1$, and $V_2$ are roughly $90$ meV, $45$ meV and $25$ meV, respectively. Note that we ignore the onsite repulsion on W sites. 
Although the strength of the interaction is also sizable ($U_W\approx 70$ meV),  the band width of W band $D_\text{W}$ is also relatively large, leading to a ratio ${U_W}/{D_\text{W}}$ is smaller than that of Mo bands. We will only consider local interactions between Mo electrons and ignore the effect of $U_\text{W}$. 
Moreover, since $U$ is large compared with other energy scales, we apply the infinite-$U$ limit, as an approximation. On the other hand,  $V_1$ and $V_2$ are treated at the mean-field level and play a crucial role by renormalizing the Mo-carriers dispersion, opening a full gap at commensurate filling, and reshaping the heavy Fermi surface in the metallic regime similarly to DMFT results~\cite{gleis2026topologicalkondoinsulatorspin}.

\section{Mean field theory}\label{sec3}
In this section, we describe the mean-field theory involving a slave boson representation and the Hartree-Fock approximation. For the on-site Coulomb interaction on Mo sites, we consider the limit where the onsite interaction $U\rightarrow\infty$ (i.e. the infinite-$U$ limit), where the doubly occupied states are projected out. 
Meanwhile, the strength of the displacement field is not strong enough to fully suppress the charge fluctuation between the singly occupied Mo states and empty Mo state, the so-called "mixed-valence" region~\cite{varmaMixedvalenceCompounds1976,lawrenceValenceFluctuationPhenomena1981}.
To study the mixed-valence physics, we represent the electron operator on the Mo sites as a product of bosonic and fermionic operators~\cite{colemanNewApproachMixedvalence1984},
\begin{equation}
f_{\bm{r},\sigma}=b_{\bm{r}}^\dagger\chi_{\bm{r},\sigma},
\end{equation}
where $b_{\bm{r}}$ is a bosonic operator, referred to as a holon, that encodes the charge degree of freedom, while $\chi_{\bm{r}\sigma}$ is a fermionic spinon operator. The physical Hilbert space is recovered by imposing the single-occupancy constraint on each Mo site, 
\begin{equation}
    b_{\bm{r}}^\dagger b_{\bm{r}}+\sum_{\sigma}\chi_{\bm{r},\sigma}^\dagger\chi_{\bm{r},\sigma}=1.
\end{equation}
At the mean-field approximation, this constraint is enforced by introducing site-dependent Lagrange multipliers $\lambda_{\bm r}$ into the free energy, which act as effective local chemical potentials for the spinons and thereby keep the corresponding $f$-electron states near the Fermi level. In addition, the local constraint imparts a $U(1)$ gauge symmetry such that our operators transform under this gauge as:
\begin{equation}\label{eq:gauge_slaveboson}
    \begin{split}
        \chi_{\bm{r},\sigma} &\rightarrow e^{i\phi (\bm r)}\chi_{\bm{r},\sigma},\\
        b_{\bm r} &\rightarrow e^{i\phi (\bm r)}b_{\bm{r}},\\
        \lambda_{\bm r} &\rightarrow \lambda_{\bm r} - i\partial_t\phi(\bm{r}).
    \end{split}
\end{equation}

Within the parton mean field theory, we can replace the physical electron operators by spinon and bosonic operators in $\hat{H}_{\text{int}}$. 
We note that the local Hubbard interaction on Mo sites is vanishing in the projected Hilbert space and, therefore, it does not appear explicitly in the projected Hamiltonian.
For  non-local interactions, we first recall the $f$-electron number operator at site $\bm r$, $\hat n_{f,\bm r}=\sum_{\sigma}f^\dagger_{\bm  r,\sigma }f_{\bm  r,\sigma}$, whose expression within the parton mean field theory  is $\hat n_{f,\bm r}=\sum_\sigma\chi_{\bm r,\sigma}^\dagger \chi_{\bm r,\sigma}$. Since both $V_1$ and $V_2$ terms involve $\hat n_{f,\bm r}$, the two interaction terms can be rewritten as
\begin{equation}
\begin{split}
     \mathcal P \hat{H}_{\text{int}} \mathcal P=V_1\sum_{\langle\bm r,\bm r'\rangle,\sigma\sigma'}
    c^\dagger_{\bm r,\sigma}c_{\bm r,\sigma}
    \chi^\dagger_{\bm r',\sigma'}\chi_{\bm r',\sigma'}+\\
    V_2\sum_{\langle\langle\bm r,\bm r'\rangle\rangle,\sigma\sigma'}
    \chi^\dagger_{\bm r,\sigma}\chi_{\bm r,\sigma}
    \chi^\dagger_{\bm r',\sigma'}\chi_{\bm r',\sigma'},
\end{split}
\end{equation}
where $\mathcal P$ projects out the double occupancies. We then perform a mean-field decoupling of the $V_1$ and $V_2$ terms by introducing expectation values $Q_{\bm{r,r'},\sigma}$ and $\Delta_{\bm{r,r'},\sigma}$. 

\begin{equation}
    \begin{split}
        &V_2\chi_{\bm{r}',\sigma}^\dagger \chi_{\bm{r}',\sigma}
        \chi_{\bm{r},\sigma}^\dagger \chi_{\bm{r},\sigma}\\&\longrightarrow 
        Q_{\bm{r,r'},\sigma} \chi_{\bm{r},\sigma}^\dagger \chi_{\bm{r}',\sigma}
        + Q_{\bm{r,r'},\sigma} ^* \chi_{\bm{r}',\sigma}^\dagger \chi_{\bm{r},\sigma}
        +\frac{|Q_{\bm{r,r'},\sigma}| ^2}{V_2},\\
          &V_1c_{\bm{r}',\sigma}^\dagger c_{\bm{r}',\sigma}
        \chi_{\bm{r},\sigma}^\dagger \chi_{\bm{r},\sigma}\\&\longrightarrow 
        \Delta_{\bm{r,r'},\sigma}  c_{\bm{r}\sigma}^\dagger \chi_{\bm{r}'\sigma}
        + \Delta_{\bm{r,r'},\sigma} ^* \chi_{\bm{r}',\sigma}^\dagger c_{\bm{r},\sigma}
        +\frac{|\Delta_{\bm{r,r'},\sigma} |^2}{V_1},
    \end{split}
    \label{eq:HS_of_int}
\end{equation}

with
\begin{equation}
    \begin{split}
          Q_{\bm{r,r'},\sigma} &= -V_2\langle \chi_{\bm{r}\sigma}^\dagger \chi_{\bm{r'}\sigma}\rangle,\\
        \Delta_{\bm{r,r'},\sigma} &= -V_1\langle c_{\bm{r}\sigma}^\dagger \chi_{\bm{r'}\sigma}\rangle.\\
    \end{split}
    \label{eq:HS_of_int2}
\end{equation}
The $\bm{r,r'}$ denote the positions of two nearby W and Mo sites or two nearby Mo sites. In our theory, we focus on the mean field decoupling that does not include spin flipping terms. 
In this manner, spin $S^z$ remains a good quantum number. 
Furthermore, we assume the variable $Q_{\bm{r,r'},\sigma}$ to be complex-valued and consider the case where time-reversal symmetry (TRS) is preserved. 
With this assumption, we have $Q_{\bm{r,r'},\uparrow} = Q^*_{\bm{r,r'},\downarrow}$ and $\Delta_{\bm{r,r'},\uparrow} = \Delta_{\bm{r,r'},\downarrow}^*$. Notice that the non-vanishing imaginary part of $Q_{\bm r,\bm r',\sigma}$ implies SLCs in our gauge. Their  feedback on  the band structure  can open an indirect topological gap, as we will show later. The above two quantities are also embedded with gauge structure. Following Eq.~\ref{eq:gauge_slaveboson}, their gauge transformation reads
\begin{equation}\label{eq:gauge_bond_fields}
    \begin{split}
        \Delta_{\bm r,\bm r',\sigma} &\rightarrow  e^{i\phi(\bm r')}\Delta_{\bm r,\bm r',\sigma},\\
        Q_{\bm r,\bm r',\sigma} &\rightarrow  e^{i(\phi(\bm r')-\phi(\bm r))}Q_{\bm r,\bm r',\sigma}.
    \end{split}
\end{equation}

Based on the above assumption of TRS, we drop the spin indices for convenience. Applying Eq.~\eqref{eq:HS_of_int} to Eq.~\eqref{eq:full_Ham}, we can write down the free energy of the resulting mean field Hamiltonian. Extremizing the free energy with respect to all variables yields the self-consistent equations. The variation with respect to $\Delta_{\bm r,\bm r'}$ and $Q_{\bm r,\bm r'}$ leads to Eq.~\eqref{eq:HS_of_int2}. The remaining three equations are given by

\begin{equation}
    \begin{split}
         b_{\bm{r}} &= -\frac{t_\text{h}}{2\lambda_{\bm r}} 
\sum_{i,\sigma}\langle c_{\bm{r+\delta_i}\sigma}^\dagger \chi_{\bm{r},\sigma}\rangle
-t_{\text{M}}b^*_{\bm r}\sum_{i,\sigma}
\langle \chi_{\bm{r},\sigma}^\dagger \chi_{\bm{r+a_i},\sigma}\rangle,\\
         1 &= \sum_{\sigma}\langle \chi_{\bm{r},\sigma}^\dagger \chi_{\bm{r},\sigma}\rangle+ |b_{\bm{r}}|^2,\\
         \nu &=\frac{1}{N}\sum_{\bm{r}\sigma}(\langle \chi_{\bm{r},\sigma}^\dagger \chi_{\bm{r},\sigma}\rangle+
         \langle c_{\bm{r},\sigma}^\dagger c_{\bm{r},\sigma}\rangle),\\
    \end{split}
    \label{eq:SC_equations}
\end{equation}
where $N$ is the number of real space lattice points and $i=1,2,3$ sums over the NN lattice vectors, $\bm{\delta}_i$. 
In the following calculation, we choose a particular gauge in which both $b_{\bm r}$ and $\Delta_{\bm r,\bm r'}$ are real. 
Details on the self-consistency equations and their solutions are provided in the Appendix \ref{appC}. 

We first analyze the bulk, translationally invariant system. Subsequently, we consider situations where translation invariance is explicitly broken by a sample edge.  In the absence of gauge fluctuations, we can choose a particular gauge such that the holon condenses to  be strictly real when solving the model. The mean field variables are treated as being spatially uniform: $Q_{\bm{r,r'},\sigma}=Q_{\sigma} =|Q|e^{is_\sigma\nu_{\bm r,\bm r'}\phi},\Delta_{\bm{r},\bm{r}'}=\Delta, b_{\bm r}=b_0, \lambda_{\bm r}=\lambda$. As a result, the mean-field Hamiltonian can be expressed in momentum space as a $2\times2$ block matrices for each spin component.

\begin{equation}\label{HMF}
    \begin{split}
         H_{\text{MF},\sigma}(\bm k) = 
        \left(
\begin{array}{cc}
    \xi_{\text{W},\bm k,\sigma} & (\Delta-b_0t_h)V_{\bm k} \\
     (\Delta-b_0t_h)V_{\bm k}^* &  \xi_{\text{M},\bm k,\sigma} 
\end{array}
        \right),
    \end{split}
\end{equation}
with 
\begin{equation}\label{HMF_notation}
    \begin{split}
        \xi_{\text{W},\bm k,\sigma} &= -2t_\text{W}\sum_{i}\cos(\bm k\cdot \bm a_i+s_\sigma\frac{2\pi}{3})-\mu+\frac{\epsilon_d}{2},\\
         \xi_{\text{M},\bm k,\sigma} &= -2\bar {t}_\text{M}
         \sum_{i}\cos\big(\bm k\cdot \bm a_i-s_\sigma\phi_\text{M}\big)-\mu-\frac{\epsilon_d}{2}+\lambda,\\
         \bar{t}_\text{M} &= \big((b_0^2t_\text{M}-\mathrm{Re}[Q])^2+(\mathrm{Im}[Q])^2\big)^{1/2},\\
        \phi_\text{M} &= \tan^{-1}\big(\frac{\mathrm{Im}[Q]}{b_0^2t_\text{M}-\mathrm{Re}[Q]}\big),\\
        V_{\bm{k}} &= \sum_{i}e^{i\bm{k}\cdot\bm{\delta}_i}.
    \end{split}
\end{equation}

One striking result in our mean field theory is that in the presence of inter-layer tunneling, the ground state is stabilized by a complex $Q$, indicating $\phi_M\neq 0$. The complex structure of $Q$ introduces an indirect gap and renders the ground state insulating for $\nu=2$.
To understand how complex $Q$ opens an indirect gap to the system, we consider a large inter-layer tunneling limit and derive the low energy effective Hamiltonian by expanding Eq.~\eqref{HMF} near two Dirac points $K$ and $K'$ of the mini BZ. 
The two Hamiltonian matrices are given by
\begin{equation}
    \begin{split}
        H_{K,\sigma} &= \left(
        \begin{array}{cc}
         E^{\text{W}}_{K\sigma}
         & v_D(-k_x+ik_y) \\
            v_D(-k_x-ik_y) &  E^{\text{M}}_{K\sigma}
        \end{array}
        \right),\\
        H_{K',\sigma} &= \left(
        \begin{array}{cc}
         E^{\text{W}}_{K'\sigma} 
         & v_D(k_x+ik_y) \\
            v_D(k_x-ik_y) &  
            E^{\text{M}}_{K'\sigma}
        \end{array}
        \right).
    \end{split}
\end{equation}
where $E^{\text{W}}_{K/K'\sigma}=-6t_\text{W}\cos(\frac{4\pi}{3}\pm s_\sigma \frac{2\pi}{3}) $ and $E^M_{K/K'\sigma}=-6\bar t_\text{M}\cos(\frac{4\pi}{3}\mp s_\sigma\phi_\text{M})$.

\begin{figure}[t!] 
\centering
\includegraphics[width = 1\linewidth]
{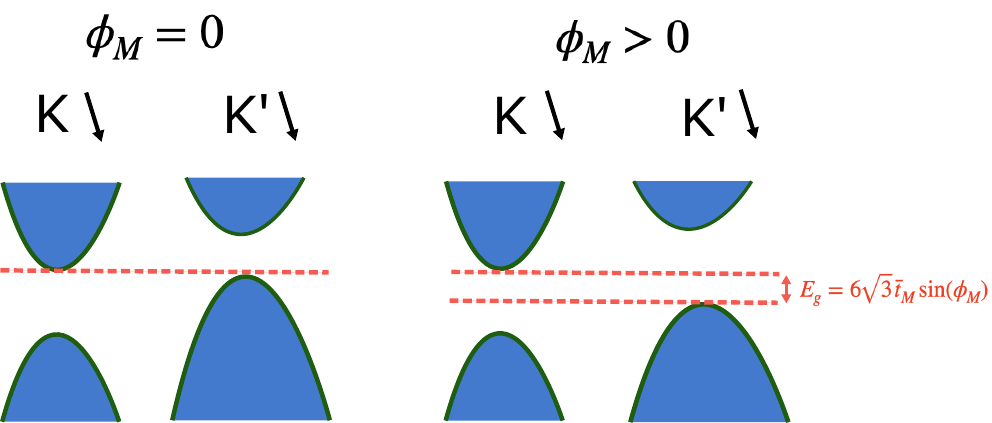}
   \caption{ Illustration for opening indirect gap. When $\phi_{\text{M}}$ is absent, no indirect gap is present since the bottom of the conduction band in the $K$ valley and the top of the valence band at the $K'$ valley are at the same energy. When $\phi_{\text{M}}$ is switched on due to the spin loop current and is positive, the bands at both $K$ and $K'$ are lifted, introducing an indirect gap.
   }\label{schematic_Eg}
\end{figure}

Here, $v_D$ is the Dirac velocity proportional to the strength of inter-layer tunneling. In addition, at filling $\nu=2$, the onsite potential terms mainly contributes to the overall energy shift and do not affect the following analysis. As an illustration, let us just focus on the spin-down sector. At Dirac points, the diagonal elements $E^{\text{W/M}}_{K/K'\sigma}$ represent the lowest energy of conduction band $E_{c,min}$ and highest energy of valence band $E_{v,max}$. We first discuss the case where $\phi_{\text{M}}=0$. For the $K$ valley, spin-up sector, we have $E^{K,\downarrow}_{c,min} = 3\bar t_\text{M}$ and $E^{K,\downarrow}_{v,max} = -6 t_\text{W}$. the $K'$ valley, spin-up sector, we have $E^{K,\downarrow}_{c,min} = 3t_\text{W}$ and $E^{K,\downarrow}_{v,max} = 3\bar t_\text{M}$. We note that $t_\text{W}$ is larger than $\bar t_{\text{M}}$ to so that $E_{c,min}$ and $E_{v,max}$ are correctly assigned. When $\phi_\text{M}$ is absent, one has $E^{K,\downarrow}_{c,min}=E^{K',\downarrow}_{v,max}=3\bar t_\text{M}$. Since the lowest point of the conduction band and the highest point of the valence band overlap, there is no indirect gap. On the other hand, if we introduce a finite $\phi_\text{M}$, $E^{K,\downarrow}_{c,min}$ and $E^{K,\downarrow}_{v,max}$ are lifted by $\pm 3\sqrt{3}\bar t_{\text{M}}\sin(\phi_{\text{M}})$. As a result, if the sign of $\phi_\text{M}$ is positive, the system hosts an indirect gap $E_g\propto \bar t_{\text{M}}\sin(\phi_{\text{M}})$. Fig.~\ref{schematic_Eg} provides a graphic illustration on how a finite $\phi_M$ can generate a non-vanishing indirect gap.

We now turn our attention to understanding the origin of the complex bond variable $Q_\sigma$.  
To gain further insight, we construct a Landau free energy involving $Q_\sigma$.
The expansion is valid under the situation where the bond order parameters develop continuously from zero and remain sufficiently small. 
This scenario can be realized when the temperature is finite and $V_2$ is sufficiently small.
Time-reversal symmetry requires $Q_{\downarrow}=Q_{\uparrow}^\ast$. 
Consequently, the free energy per particle can be expressed solely in terms of $|Q|$ and $\phi=\arg Q$.
\begin{equation}\label{eq:GLFE}
    \begin{split}
        \mathcal{F}[|Q|,\phi] &=  \kappa|Q|\cos(\phi- \theta)+\alpha|Q|^2,\\
    \end{split}
\end{equation}
where $\alpha,\kappa$ and $\theta$ are obtained by integrating out the fermionic degrees of freedom; details are provided in Appendix~\ref{appD}. 
Here, $\kappa\ge0$ is positive semidefinite and couples linearly to the bond variable $Q$. For $\alpha>0$, the free-energy saddle point satisfies $\phi-\theta=\pi$ and $|Q|=\kappa/(2\alpha)$. For $\alpha<0$, one needs to incorporate quartic order terms $\beta|Q|^4$. Since $\theta\propto(\Delta-b_0t_h)^2t_W\sin(\phi_W)\neq 0$, with $\phi_W=2\pi/3$ the Ising SOC phase in the W layer, $Q$ is complex in both cases. A complex $Q$ gives rise to a finite bond spin current, favoring the opening of a full gap and the formation of a topological Kondo insulator.

\section{Results of Mean Field Theory}\label{sec4}

\subsection{Phase Diagram}

We now describe the mean field parameters and effective band structure as we vary the filling and interactions. 
We fix the displacement field to $\epsilon_d = 3$ meV and solve the self-consistency equations as a function of  filling fraction $1.5 \leq \nu \leq 2.0$, $V_1$, and $V_2$. We observe that the main effect of $V_1$ is to modify the displacement field and hopping. Because these effects are overestimated within our mean-field approach, we use a smaller value of $V_1$. This also allows us to examine the TKSM--TKI phase boundary near $\nu = 2.0$ in greater detail.

\subsubsection{Filling $1.5< \nu<2$}

\begin{figure}[t!] 
\centering
\includegraphics[width = 1\linewidth]
{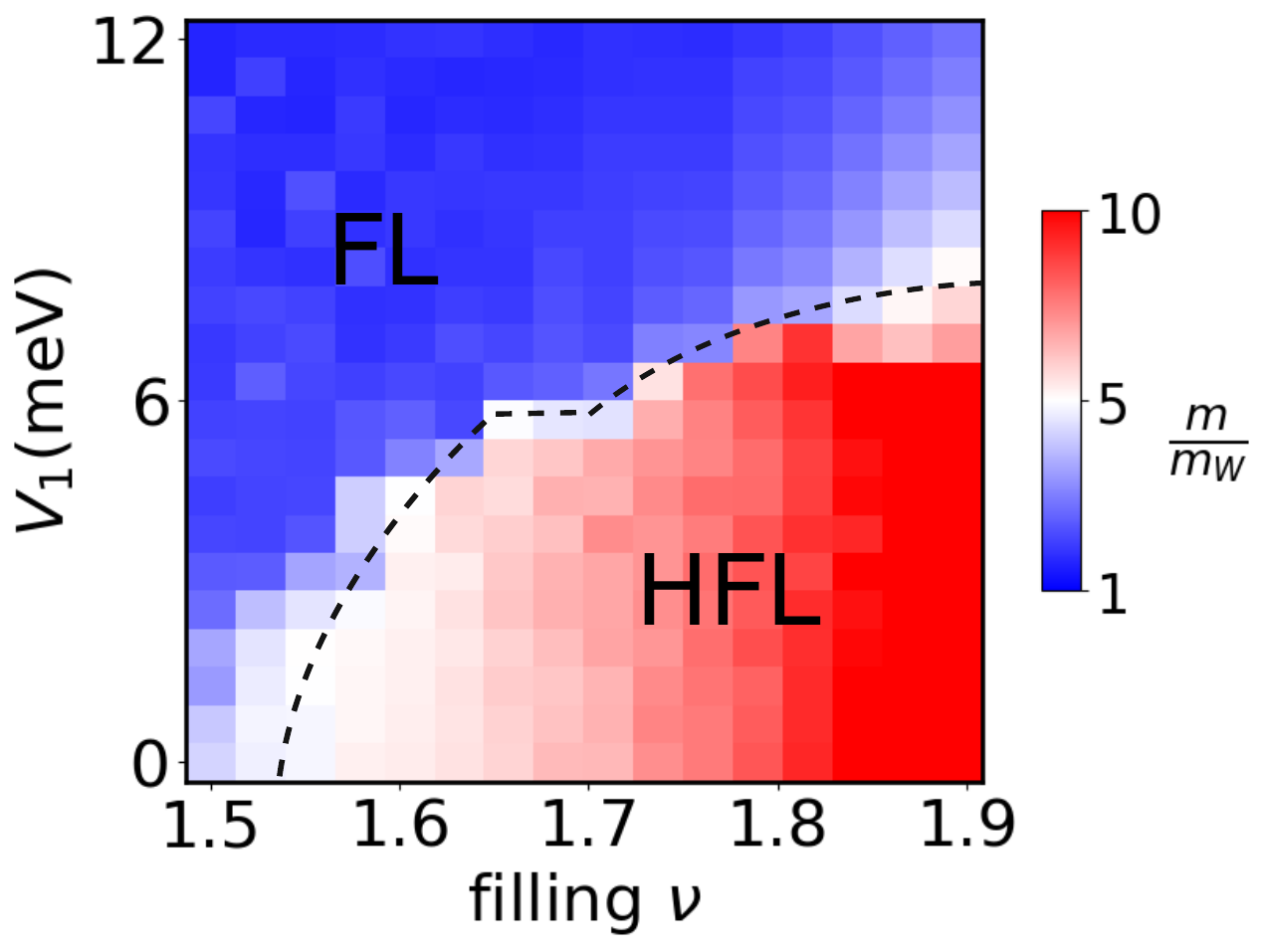}
   \caption{ The effective mass as a function of $V_1$. The plot is the colormap of the ratio between the mass with Kondo hybridization and the one of isolated WSe$_2$ band. One can see that near $\nu=2$ and weak $V_1$ region, the effective mass is large. The black dashed line marks the crossover from HFL to FL. The parameters are chosen to be $
   t_\text{W}=9~\text{meV},
   t_\text{M} = 4.5~\text{meV},
   t_\text{h}=2~\text{meV},
   \epsilon_d = 3~\text{meV}
   ,~\text{and}~V_2 = 6~\text{meV}$.
   }\label{doping_mass}
\end{figure}

\begin{figure}[t!] 
\centering
\includegraphics[width = 1\linewidth]
{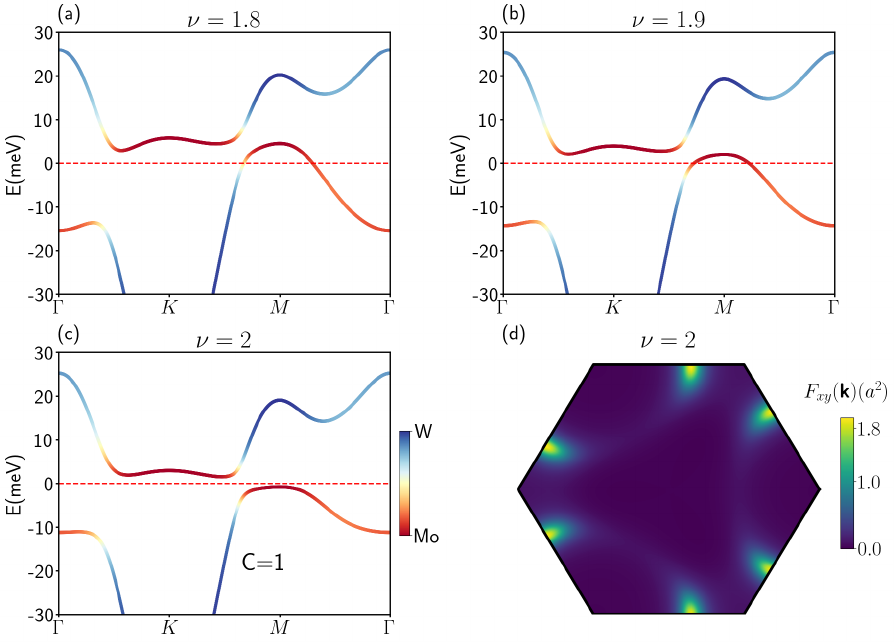}
   \caption{ The mean field band structures at different filling fraction $\nu$. Figures (a)-(c) display the band structure at fillings $\nu=1.6$, $\nu=1.8$ and $\nu=2$, respectively. The color bar corresponds to the relative occupation density of each layer. Near the gap, occupation primarily occurs on the Mo layer. As the density approaches $\nu=2$, the valence band of the FL state becomes occupied and transitions into a TKI. (d) shows the Berry curvature distribution of the band structure in (c). The momenta where the Berry curvature is prominent corresponds to the band inversion region. The value is in the unit of $a^2$,where $a$ is the moir\'e lattice constant. The parameters are chosen to be $V_1=10~\text{meV},
   V_2=6~\text{meV},
   t_\text{W}=9~\text{meV},
   t_\text{M} = 4.5~\text{meV},
   t_h=2~\text{meV},
   ~\text{and}~\epsilon_d = 3~\text{meV} 
   $. The red and blue colors in band structures (a)-(c) denote the orbital content. 
   }\label{doping_band}
\end{figure}

For $\nu<2$, we identify that the ground state is either HFL where the effective mass is large or FL state irrespective of the value of $V_2$, and therefore we focus on results in Fig.~\ref{doping_band} of $V_2=6~\text{meV}$ in this regime of filling. To distinguish the crossover from HFL phase to FL phase, we compute the effective mass $m$. To see how the mass gets larger after Kondo hybridization is switched on, we take the effective mass of isolated WSe$_2$ carriers $m_{\text{W}}$ as a reference and  evaluate the ratio $m/m_{\text{W}}$ to characterize the Fermi liquid ground state. In a two dimensional system, we can define the ratio $\frac{m}{m_\text{W}}=\frac{D(E_F)}{D_\text{W}(E_F)}$, with $D(E_F)$ and $D_W(E_F)$ being DOS of quasi-particle and isolated WSe$_2$ carriers at the Fermi energy, respectively. Fig.  \ref{doping_mass} displays the color map of $m/m_{\text{W}}$ with tuning parameters being filling fraction and $V_1$. One can identify that the effective mass is enhanced after the the Mo band hybridizes with W band. Importantly, one can distinguish the HFL and FL by defining a critical mass ratio $r_c=\frac{m}{m_\text{W}}=5$. As the effective mass of the quasi-particle is larger than $r_c$, then the ground state is a HFL, otherwise it is a FL. Furthermore, we also display mean-field band structures, as shown in Fig.~\ref{doping_band}~(a)-(c). In Fig.~\ref{doping_band} (a) and (b), the Fermi level cuts through the valence band, exhibiting a FL state. 

Once we reach a filling $\nu=2$, as shown in  Fig.~\ref{doping_band} (c), the system forms an topologically insulating ground state characterized by a finite spin-Chern number. In the following section, we will consider the affect of varying $V_2$, which influences the formation of a SLC. We will show the role the SLC plays in forming a TKI, and how to characterize its nontrivial topology.


\subsubsection{Filling $\nu=2$}

\begin{figure*} 
\centering
        \includegraphics[width = 1\linewidth]{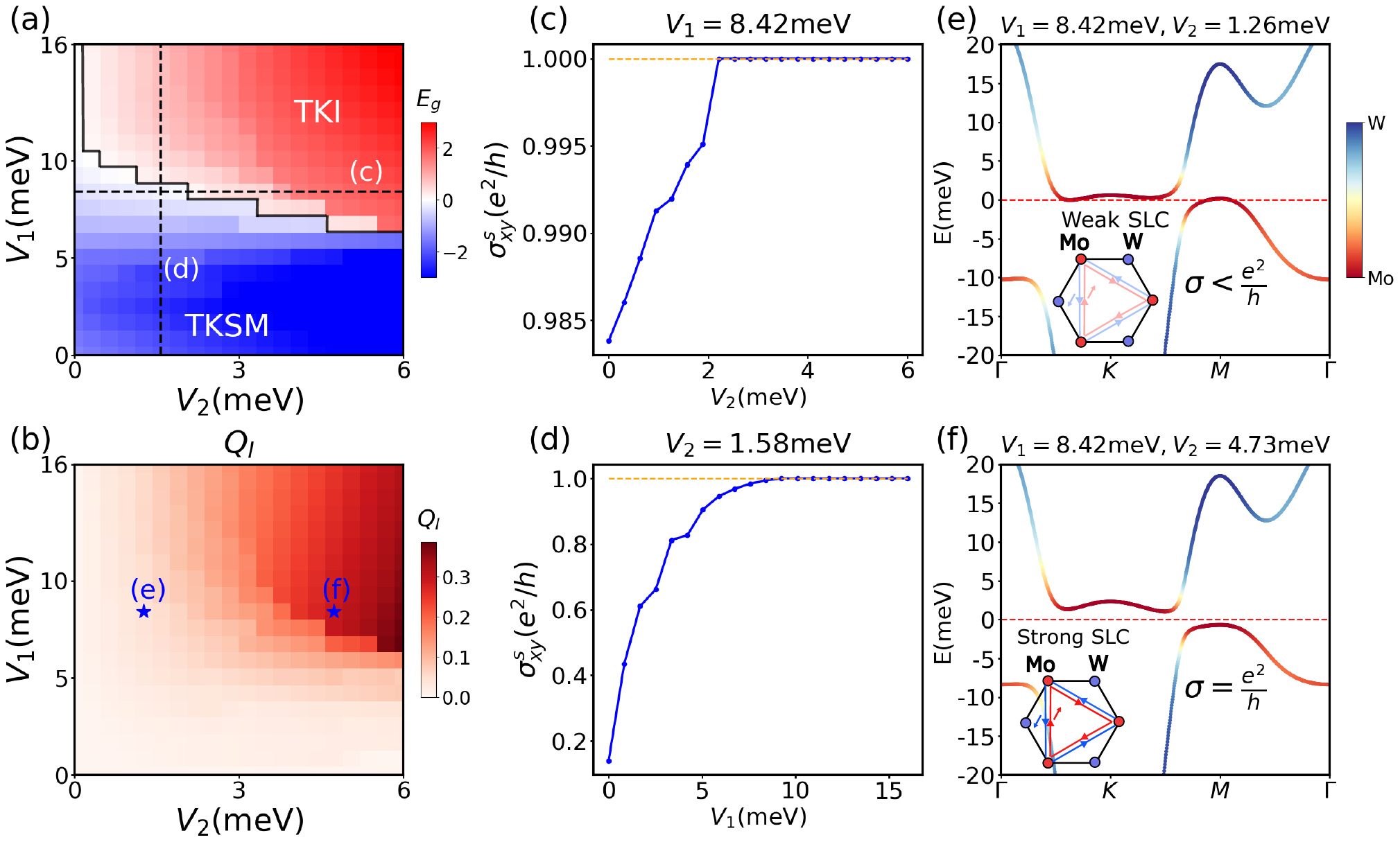}

   \caption{Phases for varying non-local interaction strengths, transport and band structures. (a) plots a color map of the full single particle gap $E_g$ as a function of $V_1$ and $V_2$ at filling fraction $\nu = 2$. 
   The blue region marks the negative $E_g$, indicating that no indirect band gap is present. The red region characterize the positive $E_g$, and it means the band structure acquires an indirect gap.
   The black solid line marks the phase boundary between topological Kondo insulator (TKI) and topological Kondo semimetal (TKSM). 
   (b) displays the color map of $Q_I$. 
   (c) (d) show the corresponding spin Hall conductance, $\sigma^s_{xy}$, for the black-dashed lines in (a). (c) plots the $\sigma^s_{xy}$  as a function of $V_2$ for $V_1= 8.42~\textrm{meV}$ fixed while (d) plots the $\sigma^s_{xy}$  as a function of $V_1$ for $V_2= 1.58~\textrm{meV}$ . (e) (f) exhibit band structure with two particular interaction strengths as marked in (b).} \label{phase_band_nu2}
\end{figure*}

In the case of $\nu = 2$, we explore the interplay between interactions and band topology to demonstrate the formation of a TKI .  For this purpose, we solve the self-consistent equations in the $V_1-V_2$ parameter space and identify the transition between  semi-metallic and insulating states. Fig.~\ref{phase_band_nu2}~(a) displays the colormap of the energy difference between the conduction band minimum and the valence band maximum $E_g$, as a function of $V_1$ and $V_2$. We start our discussion by considering increasing $V_1$ while keeping $V_2$ to be zero. At $V=0$, the system begins as a TKSM with compensated electron-hole pockets at $M$ and $K$ valleys. Increasing $V_1$ enhances $\Delta$, strengthening the inter-layer tunneling. $E_g$ also becomes less negative but never becomes positive as $V_1$ increases and does not develop an indirect band gap. Therefore, $V_1$ on its own cannot drive the system from a TKSM to TKI. On the other hand, the transition from TKSM and TKI phase is possible if we introduce the non-local interaction $V_2$ (see Fig.~\ref{doping_band}~(c)). When both $V_1$ and $V_2$ are small, the valence-band maximum still lies above the conduction-band minimum and $E_g$ is negative (see the blue-colored region in Fig.~\ref{phase_band_nu2}~(a).  If both $V_1$ and $V_2$ are sufficiently large, $E_g$ becomes positive, forming a TKI state.

Next, we examine the mean-field band structures for a fixed $V_1=8.42$ meV but with different $V_2$. In Fig.~\ref{phase_band_nu2}~(e) and (f) we show the mean field band structure along the closed path ($\Gamma\rightarrow K \rightarrow M \rightarrow \Gamma$) in the Brillouin zone. In Fig. \ref{phase_band_nu2}~(e). The system is gapless, and one can see the chemical potential cuts through both conduction and valence bands. While in Fig. \ref{phase_band_nu2} (f), the system acquires an indirect gap as an enhanced SLC (characterized by Im$[Q]$) causes a relative shift between W band and Mo band (characterized by $\arg Q$), leading to a gap opening and thereby forming an insulating ground state. 
In addition, we also scrutinize the topological nature of the ground state by computing the spin Chern number, $C^s$. 
The spin Chern number is defined as:
\begin{equation}\label{eq:total_spin_Chern}
    \begin{split}
        C^s &= \frac{1}{2}(C_{\uparrow}-C_{\downarrow}),\\
        C_{\sigma} &= \sum_{n\in\rm occ}\frac{1}{2\pi} \int_{BZ}F^{xy}_{n,\sigma}(\bm{k})dk_xdk_y.
    \end{split}
\end{equation}
Here, $C_{\sigma}$ denotes the spin-resolved Chern number. The Berry curvature, $F^{xy}_{n,\sigma}(\bm{k})$, for a given band $n$ and spin $\sigma$ is defined as
\begin{equation}\label{berry_curv}
    \begin{split}
        F^{xy}_{n,\sigma}(\bm{k}) = -2\text{Im}[\langle\partial_{k_y} u_{n,\sigma}(\bm k)|\partial_{k_x}u_{n,\sigma}(\bm k)\rangle],
    \end{split}
\end{equation}
where $|u_{n,\sigma}(\bm k)\rangle$ is the Bloch state with band index $n$, spin index $\sigma$ and momentum $\bm k$. 
The $\mathbb{Z}_2$-invariant given in Eq.~\eqref{eq:total_spin_Chern} is defined as $C^s \mod 2$ where $C^s \mod 2=1$ indicates a nontrivial topological state. 
Once the semi metallic state transitions into an insulating state, the filled valence band has $C_n^s =1$ as shown in Fig.~\ref{doping_band}~(c) (and corresponding Berry curvature distribution shown in Fig.~\ref{doping_band}~(d)). Following Eq.~\eqref{eq:total_spin_Chern}, then $C^s \mod 2=1$, and we therefore conclude that the insulating state is a TKI with nontrivial $\mathbb{Z}_2$ topological index.

A clear distinction of the topologically gapped and topologically gappless states can be made from spin Hall conductivity measurements. We compute the intrinsic contribution to the zero-temperature spin Hall conductivity, defined as 
\begin{equation}\label{SHC}
 \begin{split}
    \sigma_{SH}^{xy} 
    &=\frac{e^2}{2\pi h} \int\int_{E_{\bm k}<0} dk_xdk_y(F^{xy}_{\uparrow}(\bm k)-F^{xy}_{\downarrow}(\bm k)),
 \end{split}
\end{equation}
where $F^{xy}_{\sigma}(\bm k) = \sum_{n\in\rm occ} F^{xy}_{n,\sigma}(\bm{k})$, and $F^{xy}_{n,\sigma}(\bm{k})$ is defined in Eq.~\eqref{berry_curv}.

Fig.~\ref{phase_band_nu2}~(c) and (d) presents the zero-temperature spin Hall conductivity as a function of $V_2$. In this computation, we fix the value of $V_1$ to $8.42$ meV in Fig.~\ref{phase_band_nu2}~(c) while fix the value of $V_2$ to be $1.58$ meV.  In both plots, one can see the transition from non-quantized to quantized spin Hall conductivity. From Fig. \ref{phase_band_nu2} (c), one can see that $\sigma^s_{SH}$ is around $0.985(\frac{e^2}{h})$ at small $V_2$, whereas the spin Hall conductance saturates to $\frac{e^2}{h}$ as $V_2$ is sufficiently large (exceeding $2.21$ meV in our case). In Fig. \ref{phase_band_nu2} (d), The spin Hall conductance grows from roughly 0.2 upon increasing $V_1$ and saturate to a quantized value at $V_1 = 10$ meV. 
Once the spin Hall conductivity is quantized, we can identify this as the formation of a TKI with spin Chern number:
\begin{equation}
    C^s = \frac{h}{e^2} \sigma_{SH}^{xy}.
\end{equation}
The formation of the TKI originates from band inversion and the presence of nodal hybridization leading to a chiral Kondo exchange~\cite{guerciChiralKondoLattice2023}. The blue-red coloring in Fig.~\ref{phase_band_nu2}~(e) and~(f) indicates a jump from W-dominated orbital weight to Mo-dominated orbital weight, leading to a prominent Berry curvature at momenta where band inversion happens. Looking back to Fig.~\ref{doping_band} (d), the region where Berry curvature is relatively large corresponds to band inversion region.

\section{Edge states at and away from the Kondo insulator}\label{sec5}

We now examine boundary effects of the system.
To investigate these effects, we consider Hamiltonians constructed in real space with cylindrical boundaries (i.e. open boundary along one axis and periodic in the other). We create a lattice composed of unit-cell-preserving blocks that are repeated $N$ times along both the $x-$ and $y-$axis. For the honeycomb lattice, unit-cell-preserving terminations produce either zigzag or armchair edges. We only consider like-edges, i.e. both zigzag or armchair shown in insets of Fig.~\ref{edgeband} (a) and (c), respectively. In both scenarios, we orient the system such that the open (closed) boundary is along the $x$-axis ($y$-axis). We refer to the edges that form when terminating the $x$-direction and are extended along the $y$-direction as left and right. Under this choice of boundary termination, the zigzag edge has different sublattice termination: the left edge terminates on the Mo sites, whereas the right edge terminates on the W sites (or vice versa). For the armchair geometry, the left and right edges have identical local terminations and are therefore related under a mirror symmetry. Further details of the unit cell construction, and the corresponding Hamiltonian matrices, are provided in Appendix~\ref{appB}. For clarity, parameters $t_\text{h},~t_\text{M},~\phi_\text{M}$ mentioned in the following discussion have included the interaction effects featured by $\Delta,~b_0$, and $Q$. 

One important consequence of introducing boundaries is the modification of the mean-field self-consistency conditions. Since translational symmetry is broken, the mean-field variables generally become spatially nonuniform, which we solve using iterative techniques and simple mixing, see Appendix \ref{appB}. A detailed analysis of mean field simulations without translational symmetry in the open $x$ direction is provided in Appendix~\ref{appC}. 
As we show in  Appendix~\ref{appC}, the self-consistency solutions without translational symmetry  along the open boundary direction largely align with that of bulk scenario. As a good approximation, we can just use the bulk solution to compute edge state properties.

In what follows, we analyze how the correlation-induced SLC modifies the edge-state velocity and demonstrate how it can lead to a quantized two-point conductance. As justified in Appendix~\ref{appB} and~\ref{appC}, we use the mean field solution from periodic boundary conditions and then open the boundaries in the effective mean field Hamiltonian in the following calculations of the energy spectrum with one open boundary, the local density of states(LDOS), and two-point conductance. In particular, we focus on the self-consistency solution for the parameters $V_1=10\text{meV},V_2=6 \text{meV}$, $\nu=1.9$ and $2.0$, respectively. For the edge state velocity calculation, we do not follow the parameters from the self-consistency. Instead, we will fix $t_\text{W(M)}$ and vary $t_{\text{M(W)}}$.

\subsection{Edge band structure and local density of states}

\begin{figure*}[t]
\centering
\includegraphics[width = 1\linewidth]{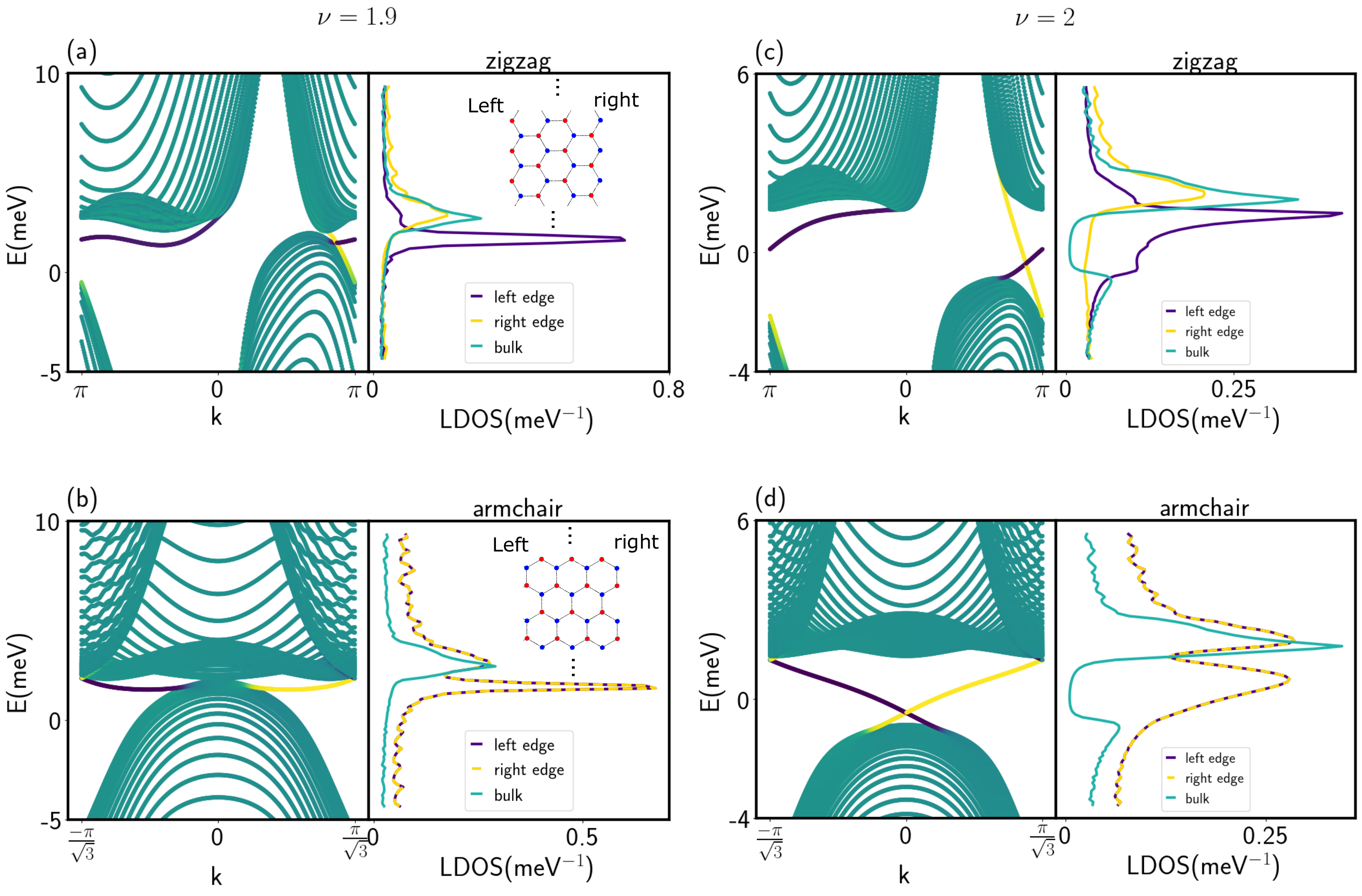}
  
\caption{Edge band structure  and the local density of state. Filling of (a) $\nu=1.9$  zigzag boundary, (b) $\nu=1.9$ armchair boundary, (c) $\nu=2$  zigzag boundary and (d) $\nu=2$ armchair boundary are plotted for band structure and corresponding local density of states (LDOS). In the band structures, the color denotes the expectation value of the position along the open boundary for the eigenstates. The purple and yellow lines indicate whether the state is localized at left or right edge, respectively, while green corresponds to the bulk, likewise for the corresponding LDOS plots. Notice that for the armchair case, the LDOS on left and right sides overlap each other, I use yellow dashed lines for the right edge to avoid the confusion. Inside the LDOS plots of (a) and (b), we provide the graphic illustration of edge geometries. }
\label{edgeband}
\end{figure*}

We begin with analyzing the spectral properties of the heterobilayer. Specifically, we analyze the band structure and the local density of states (LDOS). The LDOS is directly relevant to experiments since it can be probed by scanning tunneling microscopy (STM): the differential conductance measured by an STM tip at a bias voltage is, to leading order and at low temperature, proportional to the LDOS at energy $E$. Formally, the LDOS at energy $E$ and position $x$ is defined as
\begin{equation}\label{eq:LDOS}
\begin{split}
    D(E,x) = \sum_k -\frac{1}{\pi}\mathrm{Im}\, [g(x,k,E)],
\end{split}
\end{equation}
where $g(x,k,E)=\sum_{\alpha}\langle x,\alpha|\frac{1}{E-H(k)-i\delta}|x,\alpha\rangle$, with the index $\alpha$ denoting the sublattice (W or Mo) degrees of freedom. The symbol $\delta$ represents an infinitesimal that shifts the pole of the Green's function. In practice, we perform the numerical calculation on a finite stripe geometry $N=40$ zigzag (armchair) chains. The parameters are chosen to be ones used in Fig.~\ref{doping_band} and we consider two filling fractions: $\nu = 1.9$ and $\nu=2$.  Within the mean-field approximation, we employ the corresponding self-consistency solutions with periodic boundary conditions in both directions.

The band structures of both zigzag and armchair configurations are presented in the left panels of Fig.~\ref{edgeband}. Notice that for each state, we color-code it based on the expectation value of the position. From the spectrum, one can identify two edge modes, marked by two different colors referring to position expectation values at the two edges.  To distinguish edge and bulk contributions, we focus on three quantities: LDOS at the left edge $D_{\textrm{L}}(E) = \sum_{k}\frac{-1}{\pi}\mathrm{Im}[g(x=0,k,E)]$, right edge $D_{\textrm{R}}(E) = \sum_{k}\frac{-1}{\pi}\mathrm{Im}[g(x=N,k,E)]$, and the bulk DOS $D_{\textrm{bulk}}(E) = \sum_{k}\frac{-1}{\pi}\mathrm{Im}[g(x=N/2,k,E)]$ (see right-hand panels for Figs.~\ref{edgeband} (a)-(d)). 

\begin{figure}[t]
\centering
   \includegraphics[width =1.0 \linewidth]{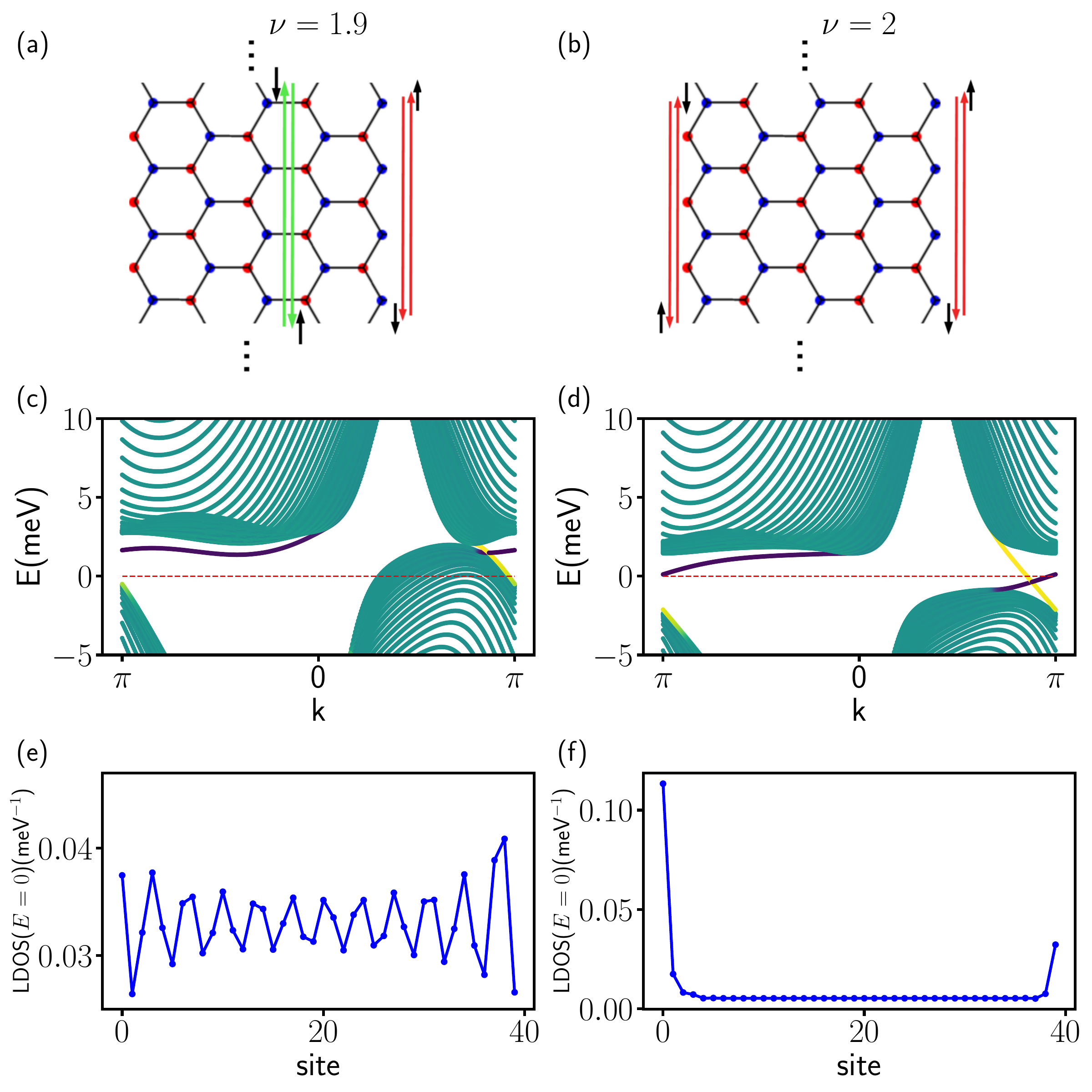}

   \caption{LDOS measurements for selected filling $\nu$.
   The column of Figures (a), (c), and (e) correspond to filling $\nu=1.9$ while (b), (d), and (f) correspond to filling $\nu=2$. (a) and (b) are illustrations of edge current distribution. The red arrows denote the edge state contribution while the green arrows denote the bulk states contribution, in (a) the velocity vanishes on the right edge as the Fermi energy does not intersect modes on that edge. 
   Notice that since the system preserves time-reversal symmetry, ($E_{\uparrow}(k) = E_{\downarrow}(-k)$) and the edge states are helical. Furthermore, the spin up edge state comes with a counter-propagating edge mode with spin down. Panels (c) and (d) represent edge band structures where the colors purple, yellow, and green correspond to the left edge, right edge, and bulk, respectively. The red lines present the position of Fermi energy. Given the position of the Fermi energy, we see that (c) exemplifies the heavy Fermi liquid  state which (d) corresponds to the topological Kondo insulating (TKI) state. (e) and (f) plot the Local density of state (LDOS) at the Fermi energy as a function of position. For the TKI state, (f), the occupation is present only at the edges. 
   }
   \label{LDOS}
\end{figure}

The right panels of Fig.~\ref{edgeband} display the density of states (DOS) for the case of zigzag and armchair boundary conditions. In the bulk, the DOS vanishes within the in-gap energy window. The width of the plateau indicates the size of the energy gap. In comparison, let us look at the LDOS at the right and left edge. Within the in-gap region, the LDOS is nonzero due to the presence of edge states. The distinction between bulk DOS and LDOS at the two edges suggests that the ground state is insulating in the bulk while remaining conducting at the edges, which is a typical signature of a topological insulator~\cite{kane2TopologicalOrder2005,mooreBirthTopologicalInsulators2010}. The spatial distribution of the edge LDOS also exhibits a strong dependence on the boundary geometry. For zigzag boundaries, the LDOS distributions at the left and right edges are inequivalent, as reflected in the distinct edge-state dispersions in Fig.~\ref{edgeband} (a) and (c). By contrast, for armchair boundaries, the LDOS distributions at the two edges are identical, consistent with the symmetric edge-state dispersions in Fig.~\ref{edgeband} (b) and (d).

We also plot the spatial profile of the local density of states in the case of $\nu=1.9$ and $\nu=2.0$, as shown in Fig.~\ref{LDOS}.  The spatial profile of the LDOS provides insight into electron transport near the Fermi surface. Fig.~\ref{LDOS} (e) corresponds to $\nu=1.9$, where nonzero LDOS is observed in both the bulk and edge regions. In particular, the LDOS exhibits a higher peak at the right edge. This feature suggests that both bulk states and the right edge state contribute to the electron transport. In comparison to the case at $\nu = 2$, shown in Fig.~\ref{LDOS} (f), the LDOS shows peaks only at the two edges; no LDOS is present in the bulk region, signifying the formation of the TKI state. Interestingly, a pronounced anisotropy in the local density of states (LDOS) is observed at the two boundaries, indicating distinct edge-state properties at opposite edges. This observation further motivates an analysis of the edge-state velocities at the two boundaries, which we present in the following subsection.

\subsection{Edge state velocity}

\begin{figure*}
\centering
    \includegraphics[width =0.9 \linewidth]{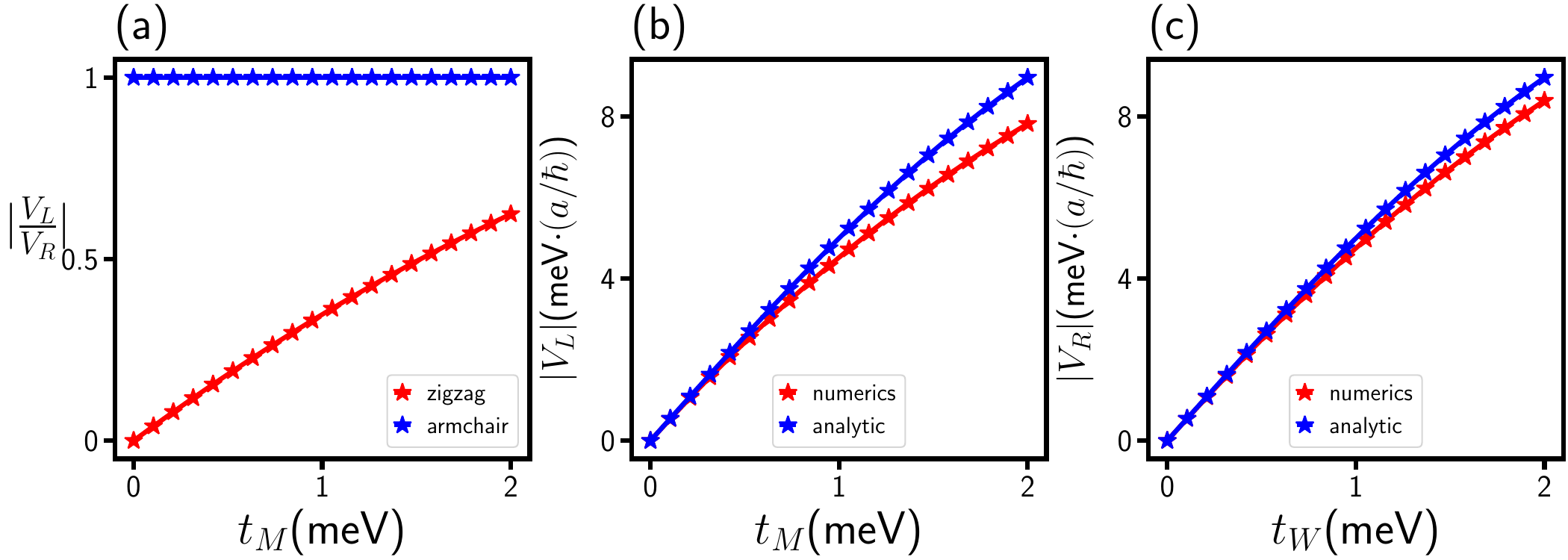}

   \caption{Edge mode velocities for varying tunneling parameter strength. The inter-layer tunneling strength is fixed to be $t_{\text{h}}=10$ (a) The ratio of edge state velocity at left and right sides as a function of $t_M$. The red curve represents the case of a zigzag boundary, while the blue curve denotes the case of an armchair boundary condition. (b) and (c) compare the analytic and numeric results of edge state velocity $V_L$ and $V_R$ in the case of the zigzag boundary condition as a function of $t_\text{M}$ and $t_\text{W}$, respectively. The unit of the velocities are represented as meV$\cdot (a/\hbar)$, with $a$ being moir\'e lattice constant .
   }\label{edgev}
\end{figure*}

Another relevant issue is the speed at which edge states propagate. To begin, we define $V_{L/R}$ as the edge state velocity at the left and right edges, respectively. In the zigzag configuration,  $V_{L/R}$ denotes the velocity at the edge where Mo (W) atoms terminate. In the following calculation, we choose $t_\text{M}$ and $t_\text{W}$ to be tuning parameters without referring to real material parameters or mean field solutions. The parameters are taken to be $ t_M = 4~\textrm{meV}$ if we vary $t_W$ and $t_W=4~\textrm{meV}$ if we vary $ t_M$. To proceed, we choose $\lambda=\epsilon_d=0$, $t_{\text{h}}=10 \text{ meV}$, and the phase factors attached to the W hopping and Mo hopping are chosen to be $\phi_W=\frac{2\pi}{3}$ and $\phi_{M}=-\frac{\pi}{3}$, respectively. To see how the edge state velocity depends on the model parameters, we compute the ratio of velocities between the two edges as a function of $ t_{\text{M}}/t_\text{W}$, as shown in Fig.~\ref{edgev}~(a). In the numerical calculation, we consider the velocity at $k=\pi$ at a zigzag edge and at $k = 0$ at an armchair edge.  In the case of the zigzag boundary condition, $\frac{V_L}{V_R}$ grows proportionally to $t_{\text{M}}/t_\text{W}$. This characteristic suggests that the anisotropy between $V_L$ and $V_R$ is controlled by the relative strength between the two hopping processes. In the case of an armchair boundary condition, the ratio remains identical at each $t_\text{M}$. In fact, the difference in group velocity between the two edges in the zigzag geometry can be understood from symmetry considerations. Under a mirror reflection with the mirror plane extending along the stripe direction, the W and Mo sites are interchanged. Since the sublattice symmetry between the W and Mo sites is explicitly broken in the system, the two edges are not related by this mirror symmetry. Consequently, there is no symmetry constraint requiring the corresponding edge states to have the same group velocity, and their velocities can therefore differ.

Now we turn to gain an analytical insight into the edge state velocity. We adopt a perturbative treatment derived in Refs.~\cite{rahmatiExplicitDerivationChiral2023,daiTwodimensionalWeylNodalline2024} for the case of the zigzag boundary condition. When the hopping parameters $t_\text{W},t_\text{M}<t_\text{h}$, the group velocity takes an approximate form (see Appendix~\ref{appE} for detailed derivation):
\begin{equation}
    \begin{split}
        V_{L/R}\approx \frac{6t_{\text{M/W}}}{\sqrt{1+4(\frac{t_{\text{M/W}}}{t_\text{h}}})^2} \sin(\phi_{\text{M/W}}),
    \end{split}
    \label{velocityLR}
\end{equation}  at $k=\pi$. This approximate relation explains the anisotropy of edge state velocity at the two edges since $t_W$ and $t_M$ are different in magnitude. In addition, Eq.~\eqref{velocityLR} also illustrates how SLC modifies the velocity since its presence renormalizes the phase $\phi_\text{M}$ such that $\phi_W$ and $\phi_M$ have different signs, featuring a counter-propagation of edge states.

One notable feature of the edge velocity is that as the $t_{\text{M(W)}}$ vanishes, the corresponding velocity $V_{t(b)}$ also vanishes. This behavior can be understood from the wavefunction structure of the localized edge state. In the absence of $t_{\text{M(W)}}$, the edge-state wavefunction has nonzero weight only on the Mo (W) sites. Since its propagation along the edge is mediated by hopping between neighboring Mo (W) sites, setting $t_{\text{M(W)}}$ eliminates this propagation channel.Consequently, the edge state remains localized rather than propagating along the edge, resulting in a dispersionless edge band and hence a vanishing group velocity.

As a comparison, we evaluate the edge state velocities both analytically and numerically in the case of zigzag boundary conditions. The results are shown in Fig.~\ref{edgev} (b) and (c). From the two plots, one can identify that the analytical results match the numerics in the small hopping strength region. As the hopping strength is tuned to be larger, the deviation between analytics and numerics becomes larger since the higher-order corrections become more significant.

\subsection{Two-point conductance}

\begin{figure}[h]
\centering
   \includegraphics[width =1\linewidth]{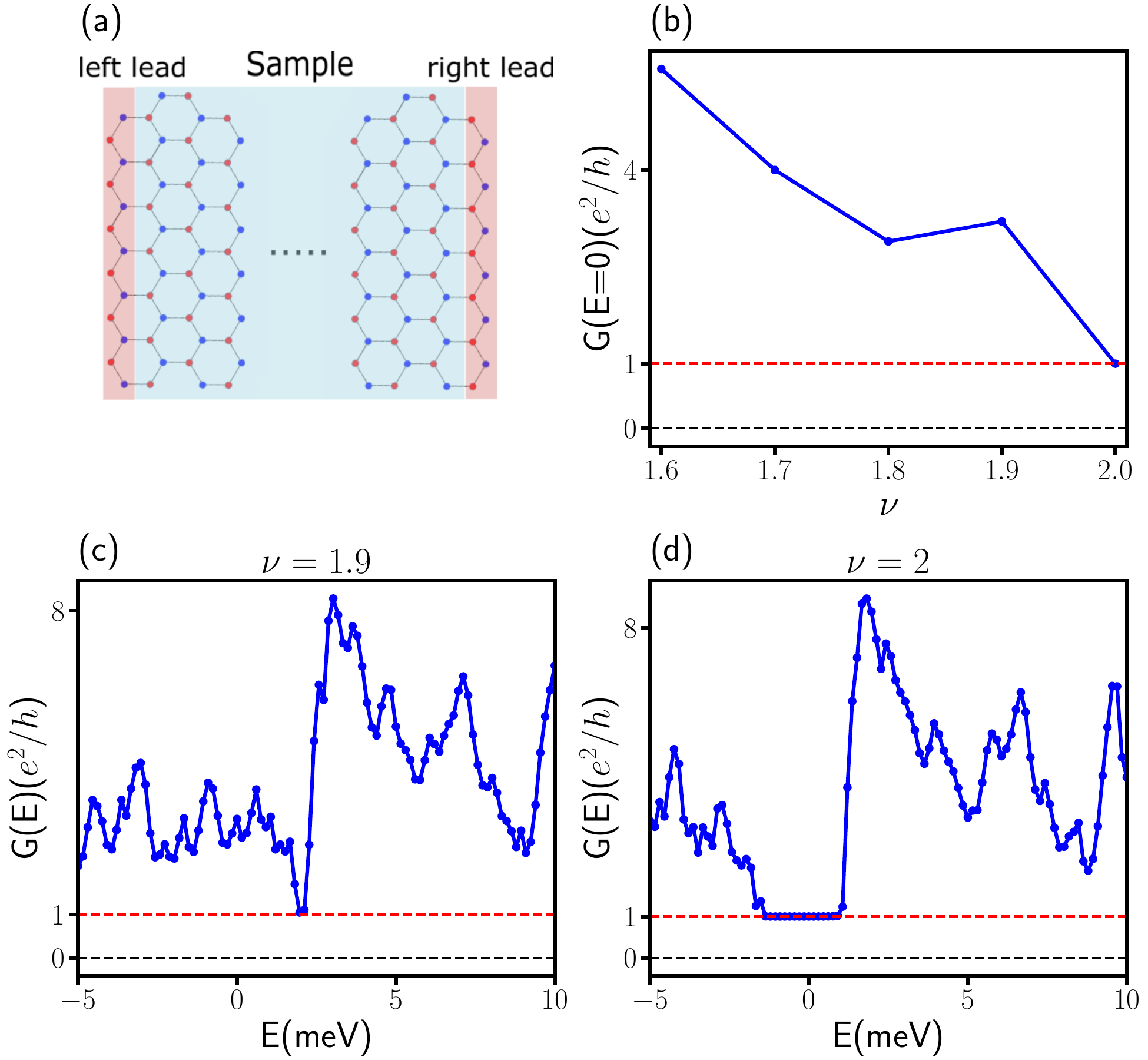}
  
   \caption{Two-point conductance for open boundary conditions in both directions. {double check this}. (a) A graphic illustration of two terminal measurements with a strip geometry. 
   The flake geometry consists of 10 zigzag chains, each of which involves 10 unit cells (with an area $10a\times\frac{20\sqrt{3}}{3}a$).  The two leads are modeled as zigzag chains as shown in the red-shaded region.
   (b) The two-point conductance as a function of  filling $\nu$ with energy fixed to the $E=0$, corresponding to Fermi level. 
   The two-point conductance as a function of  energy with filling fixed to be $\nu=$(c) $1.9$ and (d) $2$.
   }\label{Conductance}
\end{figure}

The transport quantity of primary interest is the two-point conductance, which can be measured experimentally through a multi-terminal setup~\cite{liQuantumAnomalousHall2021,Han2026}. This conductance can reveal rich physics in a topological material ~\cite{konigQuantumSpinHall2007,hasanColloquiumTopologicalInsulators2010,qiTopologicalInsulatorsSuperconductors2011,changExperimentalObservationQuantum2013} including quantized conductance arising from topological edge states. 
The conductance is evaluated using the Landauer–Büttiker formalism~\cite{georgesElectronicCorrelationsTransport1999, meirLandauerFormulaCurrent1992,Datta_2005}, which is given by
\begin{equation}\label{conductance}
    \begin{split}
        G_{\sigma}(E)= \frac{e^2}{h}\mathrm{Tr}\, [\Gamma_{L,\sigma}g_{a,\sigma}\Gamma_{R,\sigma}g_{r,\sigma}],
    \end{split}
\end{equation}
where $g_{a/r}$  denotes the advanced (retarded) Green's function, $\Gamma_{L/R,\sigma}$ is the hybridization between lead and sample on the right and left edge, respectively, and $\sigma$ corresponds to the spin component. For the real space Hamiltonian $H$ which combines non-interacting part of Eq.~\eqref{eq:full_Ham}  and real space mean fields variables, the Green function is given by

\begin{equation}
    \begin{split}
        g_{a,\sigma} &= (E-i\delta-H-(\Sigma_{L,\sigma}+\Sigma_{R,\sigma}))^{-1},\\
        \hat{g}_{r,\sigma} &= g_{a,\sigma}^*,
    \end{split}
\end{equation}
where $\pm i\delta$ denotes an infinitesimal imaginary shift that displaces the poles of the Green’s functions. The sign +(-) specifies advanced (retarded) Green function. The symbol $\Sigma_{L/R}$ is the self-energy matrix arising from integrating out the right/left lead degree of freedom. In a wide-band limit, we can assume a constant imaginary onsite energy on left/right termination. It can be expressed as

\begin{equation}\label{eq:sigma}
    \begin{split}
         \Sigma_{L/R,\sigma} &= -i\hbar/\tau\big(\sum_{\bm r\in L/R}\sum_{\alpha\in \text{W/M}} |\bm r,\sigma,\alpha\rangle
         \langle \bm r,\sigma,\alpha|\big),\\
    \end{split}
\end{equation}
where $|\bm r,\sigma,\text{W}\rangle = c^\dagger_{\bm r,\sigma}|0\rangle $ and $
         |\bm r,\sigma,\text{Mo}\rangle = \chi^\dagger_{\bm r,\sigma}|0\rangle $ are single particle states on W/Mo sites, respectively. $\tau$ characterizes the scattering rate between the lead and the sample and we take $\hbar/\tau=1 \text{ meV}$ in our calculation. $L/R$ in the summation indicates the sites located at the left/right boundary of the sample. $\Gamma_{L/R}$ can be expressed as $\Gamma_{L/R,\sigma}=  i(\Sigma_{L/R,\sigma}-\Sigma^{\dagger}_{L/R,\sigma})$. Note that the conductance, Eq.~\eqref{conductance}, is defined as the difference between the spin-up sector and spin-down sector (i.e. $G(E) = 1/2(G_{\uparrow}(E)- G_{\downarrow}(E))$).

In our numerical calculations, we consider a TMD flake with open boundaries and zigzag chain terminations, where leads are attached to the terminal arrays at opposite ends of the sample (see Fig.~\ref{Conductance}~(a) for a detailed illustration). Figure~\ref{Conductance} presents the calculated two-point conductance. We first evaluate the conductance at the Fermi energy for various filling fractions. As shown in Fig.~\ref{Conductance}~(b), for incommensurate fillings, the conductance exceeds ($e^2/h$) due to the additional contribution from bulk states. At ($\nu = 2$), however, the conductance reaches the quantized value of ($e^2/h$), originating from the presence of helical edge channels. This quantized conductance provides evidence for a topological insulating ground state.

Figures ~\ref{Conductance}~(c) and (d) show the energy dependence of the conductance for ($\nu = 1.9$) and ($\nu = 2$), respectively. For ($\nu = 1.9$), the conductance at the Fermi level exceeds ($e^2/h$), indicating a substantial contribution from bulk electronic transport. As the energy approaches the in-gap regime, the conductance decreases and reaches the quantized value of ($e^2/h$) around a bias voltage of 2 meV. In contrast, at ($\nu = 2$), Fig.~\ref{Conductance}~(d) reveals a finite energy window, approximately from -2 meV to 1 meV, where the conductance remains pinned at ($e^2/h$).

These results can be directly connected to the band structures shown in Fig.~\ref{LDOS}~(c) and (d). The energy range exhibiting quantized conductance coincides with the bulk gap region in the spectrum, where transport is dominated by the topologically protected helical edge states.

\section{Discussion and outlook}\label{sec6}

To conclude, we have theoretically investigated the AB-stacked WSe$_2$/MoTe$_2$ bilayer, extending previous investigations of valence fluctuations~\cite{guerciTopologicalKondoSemimetal2024a} by incorporating non-local interactions. Utilizing parton mean field theory and Hartree-Fock approximations of the interactions, we first show through a Ginzburg-Landau theory analysis how spin loop currents (SLC) in the Mo layer naturally arise through interlayer tunneling due to the hopping phase difference between Mo and W layer, which is a result of spin-orbit coupling. Within a self-consistent mean-field theory, we show that nonlocal interactions enhance the SLC in the Mo layer and qualitatively change its effective dispersion through the SLCs, which eventually opens a full gap at $\nu=2$.
We analyze the evolution of the HFL or FL phase from incommensurate ($1.5<\nu<1$) to commensurate ($\nu=2$) fillings as the non-local interaction strengths vary. 
The HFL and FL regimes are distinguished by the effective mass of the quasiparticles and are connected by a crossover.
As the filling decreases, the effective mass decreases accordingly, leading to a crossover from HFL to FL. 
Throughout the HFL(FL) state, a non-trivial Berry curvature is present, which leads to a finite, non-quantized spin Hall conductivity. At $\nu=2$, as $V_1$ increases, we predict a phase transition between a topological semimetalic state to a topological insulating state. This transition is a consequence of non-local interactions, which enhance inter-layer tunneling and qualitatively change the effective Mo-layer dispersion through SLCs. We derived a phase diagram showing how the two states, TKSM and TKI, can be realized by tuning $V_1$ and $V_2$.

In addition, we also discuss the edge physics of the system. For the two types of boundary conditions, zigzag and armchair boundary conditions, we compute the band structure and local density of states (LDOS). In the TKI phases, we can see two counter-propagating edge modes inside the bulk gap. At a filling slightly away from 2, the spatial profile of the LDOS displays the coexistence of both the bulk state and one of the edge states. A two-point conductance calculation based on the Landauer formalism also indicates the existence of topological edge states featured by quantized conductance. 

Given our theoretical findings, we now turn to discussion of the experimental implications. We first consider possible approaches for detecting the SLCs. One possible signature is the edge current, whose magnitude reflects the strength of the SLCs. Furthermore, rather than attempting to measure the spin current directly, a more practical approach is to apply an out-of-plane Zeeman field to induce a spin imbalance, thereby converting the spin current into a net charge current. The resulting charge current can then be detected through a standard conductance measurement.

Imaging techniques may also provide a complementary route to probing the presence of SLCs  in the bulk. 
In particular, nano-SQUID magnetometry~\cite{Jos_Mart_nez_P_rez_2017} could potentially detect local magnetic signatures associated with circulating charge currents induced by a spin-imbalance, which can be achieved by applying a Zeeman field or proximity to a magnetic substrate. Although the spatial resolution of nano-SQUID magnetometry may be insufficient to resolve individual moir\'{e} lattice sites, separated by approximately $5\,$nm, this technique could potentially image larger spatial domains associated with SLCs, provided that they generate a detectable magnetic signal.

The SLC mechanism can be experimentally verified by tuning the strength of non-local interaction $V_1$ and $V_2$, and measuring its effect on quantities that are sensitive to the gap, such as the temperature-dependence of the resistivity. In practice, the thickness of dielectric layers determines the magnitude of $V_1$ and $V_2$. Upon decreasing the thickness, the screening effect gets more pronounced and thereby lowering $V_1$ and $V_2$. Furthermore, we expect that if the transition between the insulating and semi-metallic state originates from the SLC mechanism, \agx{de}\ag{in}creasing thickness can drive a transition from TKSM to TKI.

Lastly, we comment on our results on AB-stacked hetero-bilayers in relation to recent experimental studies. 
While earlier experimental work on heavy-fermion physics in MoTe${}_2$/WSe${}_2$ hetero-bilayers were originally cited as AB-stacked~\cite{zhaoGatetunableHeavyFermions2023}, a follow-up study suggested that the stacking order in these studies was not identified correctly~\cite{han2026stackingorderdependentelectronicpropertiesmote2wse2}.
According to that follow-up study, heavy-fermion physics, including recent evidence of a TKI at filling $\nu = 2$~\cite{han2025evidencetopologicalkondoinsulating} ,
arises in AA-stacked devices but is absent in AB-stacked ones~\cite{han2026stackingorderdependentelectronicpropertiesmote2wse2}.

Nevertheless, our theoretical results on AB-stacked MoTe${}_2$/WSe${}_2$ compares favorably with the experimental phase diagram of AA-stacked MoTe${}_2$/WSe${}_2$.
Based on preliminary theoretical results on the latter, we believe that the reason for favorable comparison to experiments with different stacking order is that AB-stacked and AA-stacked MoTe${}_2$/WSe${}_2$ are governed by a conceptually similar Hamiltonian, with parameters that depend on the stacking order. Presumably, our theory here overestimates the interlayer tunneling present in the experiments on AB-stacked MoTe${}_2$/WSe${}_2$ and is closer to that of AA-stacked MoTe${}_2$/WSe${}_2$. It would be interesting to see whether enhancing the interlayer tunneling experimentally, e.g.\ by means of pressure, will eventually give rise to heavy-fermion and TKI physics in AB-stacked devices. As a future exploration, we can extend our theory to the AA-stacked case and we believe that the core conceptual findings from AB-stacked MoTe${}_2$/WSe${}_2$ will carry over.

\section*{Acknowledgments}
We thank Andrei Bernevig, Piers Coleman,  Kin-Fai Mak, and Jie Shan  for useful discussions. PJ was supported by Office of Basic
Energy Sciences, Material Sciences and Engineering Division, U.S. Department of Energy (DOE) under Contract
DE-FG02-99ER45790. AG acknowledges support from the Abrahams Postdoctoral Fellowship of the Center for Materials Theory at Rutgers University. K.P.L. and J.H.P. has been supported in part by the US Department of Energy Office of Science, Basic Energy Sciences, under Award No.~DE-SC0026023 (work involved the conceptualization and construction of the project, as well as the interpretation of results and the presentation).The work of AJM was supported in part by the National Science Foundation (NSF) Materials
Research Science and Engineering Centers (MRSEC) program
through Columbia University under the Precision-Assembled
Quantum Materials (PAQM) Grant No. DMR-2011738. 
The Flatiron Institute is a division of the Simons Foundation.

\onecolumngrid

\appendix
\counterwithin*{equation}{section} 
\renewcommand\theequation{\thesection\arabic{equation}} 

\section{The Hamiltonian in the zigzag and armchair open boundary}\label{appB}
\begin{figure}[h] 
\centering
\includegraphics[width = 1\linewidth]{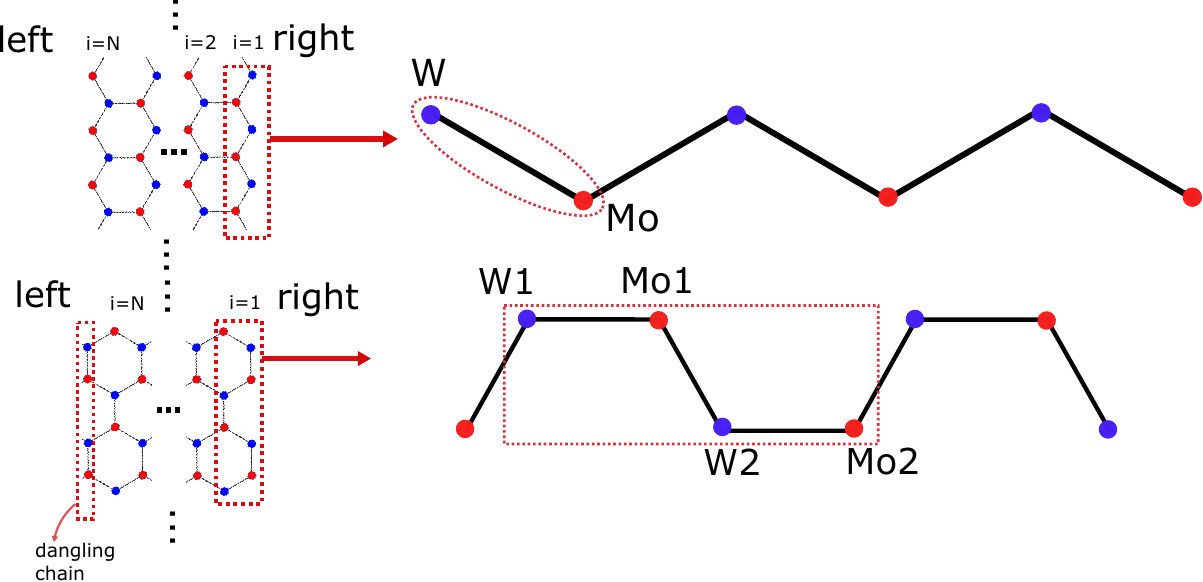}
   \caption{Schematic illustration
of the composition of zigzag(armchair) stripe. In the zigzag configuration, the geometry is made of zigzag chains whose periodicity is captured by a unit cell containing two atoms. In the armchair configuration, the geometry consists of armchair chains with four atoms per unit cell and one dangling chain.}\label{chain_OBC}
\end{figure}
For honeycomb lattices, such as this one, there are two types of boundary conditions: zigzag and armchair, which we treat as a composition of $N$ chains (see Fig. ~\ref{chain_OBC}). In the case of the zigzag configuration, the unit-cell inside a zigzag chain contains two sublattices, leading to a $2N\times2N$ Hamiltonian matrix. On the other hand, an armchair flake consists of chains whose unit cell involves four atoms. Furthermore, to make the two edge symmetric, one needs to attach an additional chain whose uni-cell contains two atoms. As a result the size of the Hamiltonian is $(4N+2)\times (4N+2)$. We start by writing down the Hamiltonian in the case of the zigzag boundary condition, which is given by

\begin{equation}\label{Hz1}
    \begin{split}
        H_{\text{zigzag}}(k) = \left (
        \begin{array}{cccc}
           H_{0,z}(k)  & H_{1,z}(k) && \\
          H^{\dagger}_{1,z}(k)& H_{0,z}(k)&\ddots&\\
           &\ddots&\ddots  & H_{1,z}(k)\\
           &&H^{\dagger}_{1,z}(k)& H_{0,z}(k)
        \end{array}
        \right),
    \end{split}
\end{equation}
with 
\begin{equation}\label{Hz2}
    \begin{split}
        H_{0,z} (k) &= 
        \left(
        \begin{array}{cc}
           -2t_\text{W}\cos(k+\phi_\text{W})+\frac{\epsilon_d}{2}-\mu&-2t_{\text{h}}\cos\frac{k}{2}\\
           -2t_{\text{h}}\cos\frac{k}{2}& -2t_\text{M}\cos(k+\phi_\text{M})-\frac{\epsilon_d}{2}-\mu+\lambda
        \end{array}
        \right),\\
          H_{1,\text{z}} (k) &= 
        \left(
        \begin{array}{cc}
           -2t_\text{W}\cos(\frac{k}{2}-\phi_\text{W})&-t_{\text{h}}\\
           0& -2t_\text{M}\cos(\frac{k}{2}-\phi_\text{M})
        \end{array}
        \right).\\
    \end{split}
\end{equation}
Subsequently, the Hamiltonian with the armchair boundary condition is given by
\begin{equation}
    \begin{split}
        H_{\text{armchair}}(k) = \left (
        \begin{array}{ccccc}
           H_{0,a}(k)  & H_{1,a}(k) && \\
          H^{\dagger}_{1,a}(k)& H_{0,a}(k)&\ddots&\\
           &\ddots&\ddots  & H_{1,a}(k)\\
           &&H^{\dagger}_{1,a}(k)& H_{0,a}(k)&H_{2,a}(k)\\
           &&&H^\dagger_{2,a}(k)&\bar H_{0,a}(k)
        \end{array}
        \right),
    \end{split}
\end{equation}
with $H_{0,a}(k),\bar H_{0,a}(k),H_{1,a}(k)$ and $H_{2,a}(k)$ defined as

\begin{equation}
    \begin{split}
        H_{0,a} (k) &= 
        \left(
        \begin{array}{cccc}
            \frac{\epsilon_d}{2}-\mu & t_h  &-t_\text{W}(1+e^{i\sqrt{3}k})e^{i\phi_\text{W}}& t_h e^{i\sqrt{3}k} \\
             & -\frac{\epsilon_d}{2}-\mu-\lambda &t_h& -t_\text{M}(1+e^{i\sqrt{3}k})e^{i\phi_\text{M}}\\
          &&\frac{\epsilon_d}{2}-\mu &t_\text{h}\\
          &H.c&&-\frac{\epsilon_d}{2}-\mu-\lambda
        \end{array}
        \right),\\
        \bar H_{0,a} (k) &= 
        \left(
        \begin{array}{cc}
          -\frac{\epsilon_d}{2}-\mu-\lambda  & t_h  \\
            0 &  \frac{\epsilon_d}{2}-\mu \\
        \end{array}
        \right)
        \\
         H_{1,a} (k) &= 
        \left(
        \begin{array}{cccc}
           -t_\text{W} e^{i\phi_\text{W}} &t_\text{h}&-t_\text{W}(1+e^{i\sqrt{3}k})e^{-i\phi_\text{W}}&0\\
            0&-t_\text{M} e^{-i\phi_\text{M}}&0& 0 \\
           0&0&-t_\text{W}e^{i\phi_\text{W}}&0\\
              0& -t_\text{M}(1+e^{-i\sqrt{3}k})e^{i\phi_\text{M}}&t_\text{h}&-t_\text{M} e^{-i\phi_\text{M}}
        \end{array}
        \right),\\
         H_{1,a} (k) &= 
        \left(
        \begin{array}{cccc}
           t_\text{h}&-t_\text{W}(1+e^{i\sqrt{3}k})e^{-i\phi_\text{W}}\\
            -t_\text{M} e^{-i\phi_\text{M}}&0 \\
           0&-t_\text{W}e^{i\phi_\text{W}}\\
               -t_\text{M}(1+e^{-i\sqrt{3}k})e^{i\phi_\text{M}}&t_\text{h}.
        \end{array}
        \right)
    \end{split}
\end{equation}

\section{Details of the self-consistent equation}\label{appC}

\subsection{Spatially uniform variables}

In this section, we provide some technical details about our mean-field theory. Recall the mean field Hamiltonian (i.e. Eq.~\eqref{HMF}) in the presence of translation symmetry in both $x$ and $y$ direction and express the Hamiltonian matrix in terms of Pauli matrices.
\begin{equation}
    \begin{split}
        H_{\text{MF},\sigma}(\bm k) = d_{0,\bm k,\sigma}\mathbb{I} + \bm d_{\bm k,\sigma}\cdot \bm\tau,
    \end{split}
\end{equation}
where $\mathbb{I}$ is a two-by-two identity matrix, and $\bm \tau=(\tau_1,\tau_2,\tau_3)$ denotes Pauli matrices acting in layer space. The expressions of $d_{0,\bm k,\sigma}$ and the vector $\bm d_{\bm k,\sigma}=(d_{1,\bm k,\sigma},d_{2,\bm k,\sigma},d_{3,\bm k,\sigma})$ are given by
\begin{equation}\label{d_expression}
    \begin{split}
        d_{0,\bm{k},\sigma} &= \frac{\xi_{\text{W},\bm k,\sigma}+\xi_{\text{M},\bm k,\sigma}}{2},\\
        d_{1,\bm{k},\sigma} &= (\Delta+b_0t_\text{h})\mathrm{Re}(V_{\bm k}),\\
         d_{2,\bm{k},\sigma} &= (\Delta+b_0t_\text{h})\mathrm{Im}(V_{\bm k}),\\
           d_{3,\bm{k},\sigma} &= \frac{\xi_{\text{W},\bm k,\sigma}-\xi_{\text{M},\bm k,\sigma}}{2}.
    \end{split}
\end{equation}
The energy eigenvalue can be obtained analytically and expressed as
\begin{equation}\label{spectrum_d}
\begin{split}
    E_{\pm,k,\sigma} =d_{0,\bm{k},\sigma}
    \pm|\bm{d}_{\bm{k},\sigma}|,\\
\end{split}
\end{equation}

with $|\bm{d}_{\bm{k},\sigma}|=\sqrt{d_{1,\bm{k},\sigma}^2+d_{2,\bm{k},\sigma}^2+d_{3,\bm{k},\sigma}^2}$. Then the free energy is given by 
\begin{equation}\label{free_energy}
    \begin{split}
        \mathcal{F} = -k_BT\frac{1}{N_{\bm k}}\sum_{\bm{k},n=\pm,\sigma}\ln (1+e^{-\beta E_{n,\bm{k},\sigma}})+\frac{3|\Delta|^2}{V_1}
        +\frac{3\sum_{\sigma}|Q_\sigma|^2}{V_2}+\mu \nu+ \lambda(b_0^2-1),
    \end{split}
\end{equation}
where $N_{\bm{k}}$ is the number of points in the $\bm{k}$-mesh. The self-consistent equations can be obtained by requiring the variation of free energy with respect to all variables to be zero: $\frac{\partial \mathcal{F}}{\partial Q_\sigma}=\frac{\partial \mathcal{F}}{\partial \Delta}=\frac{\partial \mathcal{F}}{\partial b_0}=\frac{\partial \mathcal{F}}{\partial \lambda}=\frac{\partial \mathcal{F}}{\partial \mu} = 0$. The equations are given as follows.   
\begin{equation}\label{MFeqs_bulk}
    \begin{split}
        \frac{\partial \mathcal{F}}{\partial Q_\sigma}& = \frac{1}{N_{\bm k}}\sum_{k}\big(\frac{1}{2}(f(E_{+,\bm{k},\sigma})+f(E_{-,\bm{k},\sigma}))+\frac{d_{3,\bm{k},\sigma}}{2|\bm{d}|}(f(E_{-,\bm{k},\sigma})-f(E_{+,\bm{k},\sigma}))\big)(F_c(\bm k)+iF_s(\bm k))+\frac{12Q_\sigma^*}{V_2}=0,\\
         \frac{\partial \mathcal{F}}{\partial \Delta} &=\frac{1}{N_{\bm k}}\sum_{\bm{k},\sigma}\frac{(\Delta+b_0t_\text{h})|V_{\bm{k}}|^2}{|\bm{d}|}
         (f(E_{+,\bm{k},\sigma})-f(E_{-,\bm{k},\sigma}))+\frac{6\Delta}{V_1}=0\\
        \frac{\partial \mathcal{F}}{\partial b_0} &= \frac{1}{N_{\bm k}}\sum_{\bm{k},\sigma}\big[(-t_\text{M}b_0F_c)(f(E_{+,\bm{k},\sigma})+f(E_{-,\bm{k},\sigma}))+\\
        &\big(\frac{2b_0(\Delta+rt_\text{h})|V_{\bm{k}}|^2}{|\bm{d}|}-\frac{d_{3,\bm{k},\sigma}}{|\bm{d}|}\big)(f(E_{+,\bm{k},\sigma})-f(E_{-,\bm{k},\sigma}))\big]
        +2\lambda r=0,\\
        \frac{\partial \mathcal{F}}{\partial \lambda} &= 
        \frac{1}{N_{\bm k}}\sum_{\bm{k},\sigma}\frac{1}{2}(f(E_{+,\bm{k},\sigma})+f(E_{-,\bm{k},\sigma}))+\frac{d_{3,\bm{k},\sigma}}{2|\bm{d}|}(f(E_{-,\bm{k},\sigma})-f(E_{+,\bm{k},\sigma}))+b_0^2-1=0,\\
        \frac{\partial \mathcal{F}}{\partial \mu} &= \frac{1}{N_{\bm k}}\sum_{\bm{k},\sigma}
        (f(E_{+,\bm{k},\sigma})+f(E_{-,\bm{k},\sigma}))-\nu=0.
    \end{split}
\end{equation}
 Here, we have defined short notations $F_c(\bm k) = 2\sum_{i}\cos(\bm k\cdot \bm a_i)$ and $F_s(\bm k) = 2\sum_{i}\sin(\bm k\cdot \bm a_i)$ for the form factors. $f(E)$ is the Fermi function.

\begin{figure*}[t]
\centering
\includegraphics[width = 0.9\linewidth]{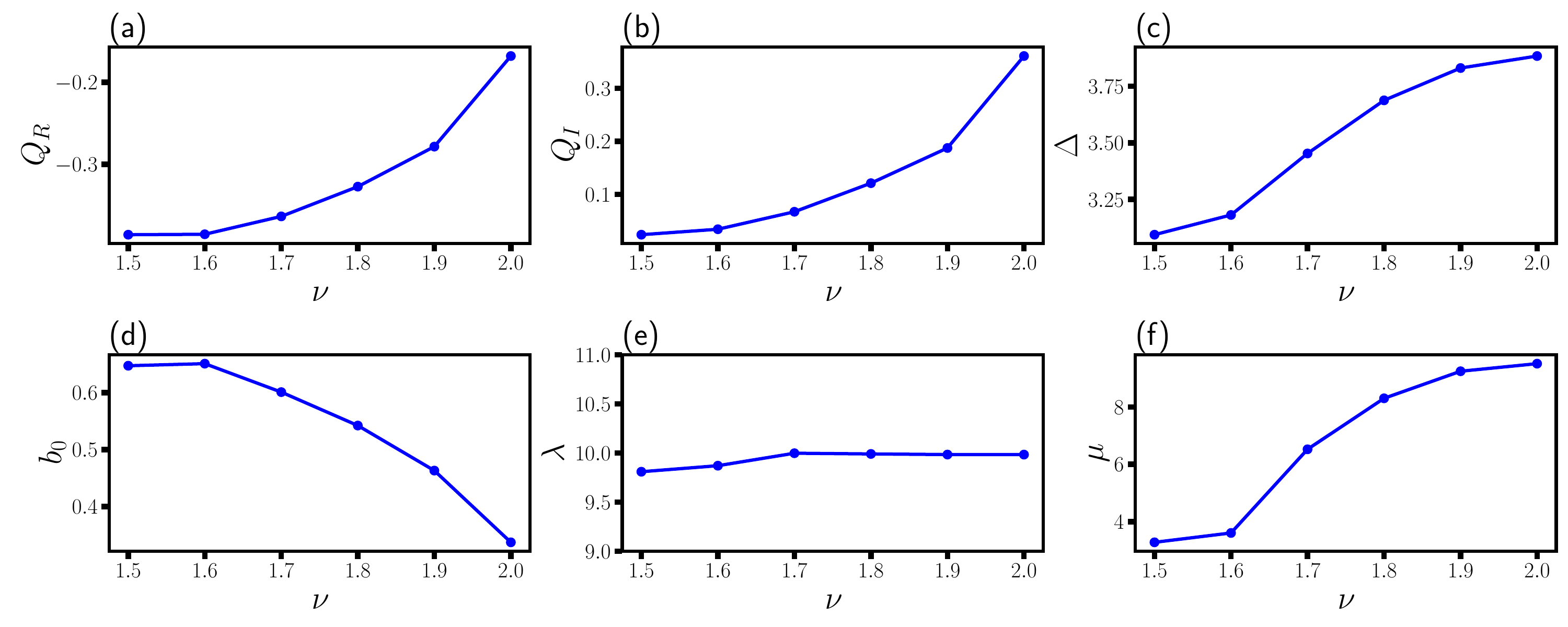}
   \caption{Self-consistent solutions of mean field equations as a function of filling fraction $\nu$. Figures (a)-(f) show the self-consistently determined parameters of Eqs.~\eqref{eq:HS_of_int2} and~\eqref{eq:SC_equations} between filling $\nu=1.5$ to $2.0$.  We choose the parameters $V_1=15~\text{meV},
   V_2=10~\text{meV},
   t_\text{W}=9~\text{meV},
   t_\text{M} = 4.5~\text{meV},
   t_h=2~\text{meV},
   \epsilon_d = 3~\text{meV}$.}\label{sol_doping}
\end{figure*}
Figure~\ref{sol_doping} presents self-consistency solutions as a function of filling fractions. One obvious feature is that the interaction-induced SLC is substantially weaken as the filling is further away from 2. In contrast, the boson condensation $b_0$ is enhanced as $\nu$ goes down, indicating a gradual enhancement of the itinerant character of the electrons upon electron doping.

\subsection{The zigzag boundary condition}

\begin{figure}[h]
\centering
   \includegraphics[width=0.7\linewidth]{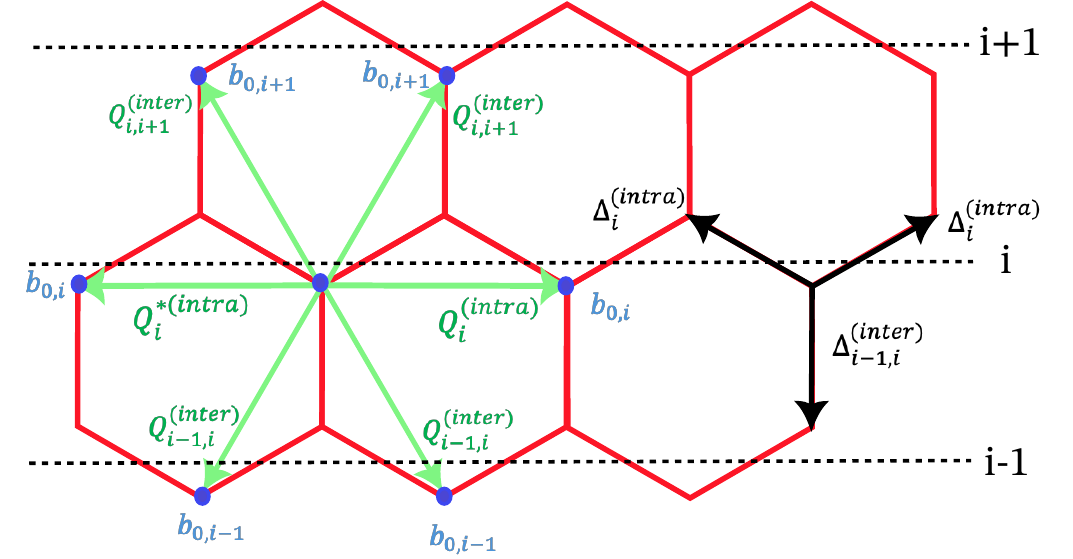}
   \caption{The graphic illustration of mean field variables in the zigzag boundary condition. }\label{MF_order_zigzag}
\end{figure}

Now, we present the mean-field formulism in the case of the zigzag boundary condition. Upon introducing the boundary condition, the translation symmetry is broken in one direction (here we assume it is the $x$ direction) and W and Mo carriers can be denoted by fermionic operators $c_{i,k,\sigma}$ and $\chi_{i,k,\sigma}$, respectively. Here $i$ denotes the position along the $x$ direction and $k$ is the momentum along the $y$ direction. Notably, introducing the boundary condition also breaks $C_{3z}$ symmetry, which changes the way of defining bond variables $\Delta$  and $Q$. As shown in Fig. \ref{MF_order_zigzag}, we define bonds connecting sites within the same zigzag chain as $\Delta^{\text{(intra)}}$ and $Q_\sigma^{\text{(intra)}}$. While the bonds connecting two sites on neighboring zigzag chains are defined as $\Delta^{\text{(inter)}}$  and $Q_\sigma^{\text{(inter)}}$. In our calculation, we assume that the number of zigzag chains is $N_c$ and the number of points in the k-mesh along the translation invariant direction is $N_k$. Since the translation symmetry is broken along the $x$ direction, the mean field variables should be position-dependent and acquire a spatial profile in the self-consistency.

In zigzag configuration, the mean-field Hamiltonian $H_{MF,\sigma}(\bm{k})$ takes the form similar to Eq.~\eqref{Hz1} and Eq.~\eqref{Hz2}. The corresponding free energy is given by:
\begin{equation}\label{free_energy_obc}
    \begin{split}
        \mathcal{F} =& -k_BT\frac{1}{N_kN_c}\sum_{k,\sigma} \text{Tr}[\text{ln}(1+e^{-\beta H_{MF,\sigma}(k)})] \\
        &+\frac{1}{N_kN_c}\sum_{i,\sigma}\big( \frac{2|Q_{i,\sigma}^{(\text{intra})}|^2}{V_2}
        +\frac{2 |Q^{(\text{inter})}_{i,i+1,\sigma}|^2}{V_2}
        + \frac{2\Delta^{(\text{intra})2}_{i}}{V_1}
        + \frac{\Delta^{(\text{inter})2}_{i,i+1}}{V_1}\big)\\
        &+\frac{1}{N_kN_c}\sum_i\lambda_i(b_{0,i}^2-1)
        +\mu \nu.
    \end{split}
\end{equation}

Minimizing the free energy, the self-consistency equations are given by: 

\begin{equation}\label{MFeqs_obc1}
    \begin{split}
        Q^{\text{(intra)}}_{i,\sigma} &= \frac{V_2}{2N_k}\sum_k
        \big(\cos(k)-i\sin(k)\big)\langle 
\chi^\dagger_{i,k,\sigma}\chi_{i,k,\sigma}\rangle,\\
        Q^{\text{(inter)}}_{i,i+1,\sigma} &= \frac{V_2}{2N_k}\sum_k
         \big(\cos(k/2)+i\sin(k/2)\big)\langle \chi^\dagger_{i+1,k,\sigma}\chi_{i,k,\sigma}\rangle.\\
    \end{split}
\end{equation}
Here, the variable is defined for the spin-up sectors. The spin-down counterpart can be obtained by applying complex conjugation, as mentioned in the main text. To proceed, we write down the self-consistency equation for boson condensation $b_0$
\begin{equation}\label{MFeqs_obc2}
    \begin{split}
          &2\lambda b_{0,i} 
         + \frac{1}{N_k}\sum_{k,\sigma}\big( 4t_\text{h} \cos(k/2) \mathrm{Re}[\langle c^{\dagger}_{i,k,\sigma}\chi_{i,k,\sigma} \rangle]
         +2t_\text{h}  \mathrm{Re}[\langle c^{\dagger}_{i+1,k,\sigma}\chi_{i,k,\sigma} \rangle]\\
         &
        +\frac{8}{V_2}b_{0,i+1}t_{M}Q^{\text{(inter)}}_{R,i+1,i,\sigma} 
        +\frac{8}{V_2}b_{0,i}t_{M}Q^{\text{(intra)}}_{R,i,\sigma} 
        +\frac{8}{V_2}b_{0,i-1}t_{M}Q^{\text{(inter)}}_{R,i,i-1,\sigma}\big)=0. 
\end{split}
\end{equation}
Here, we have introduced real part $Q_R$ and and imaginary part $Q_I$. Note that we have substituted the terms related to $\langle \chi^\dagger_{i,k,\sigma}\chi_{i,k,\sigma}\rangle$ by $Q^{\text{(intra)}}_{R,i}$ or $Q^{\text{(inter)}}_{R,i,i+1}$. The mean field equation for variables $ \Delta^{\text{(intra)}}_{i}$ and $\Delta^{\text{(inter)}}_{i,i+1}$ can be expressed as

\begin{equation}\label{MFeqs_obc3}
    \begin{split}
        \Delta^{\text{(intra)}}_{i} &= -\frac{V_1}{N_k}\sum_{k,\sigma}
        \cos(k/2)\mathrm{Re}[\langle c^{\dagger}_{i+1,k,\sigma}\chi_{i,k,\sigma} \rangle],\\
         \Delta^{\text{(inter)}}_{i,i+1} &= -\frac{V_1}{N_k}\sum_{k,\sigma}
        \mathrm{Re}[\langle c^{\dagger}_{i+1,k,\sigma}\chi_{i,k,\sigma} \rangle].
    \end{split}
\end{equation}
The other two self-consistent equations arise from fixing the total number of electrons and the local bosonic constraint:
\begin{equation}\label{MFeqs_obc4}
    \begin{split}
        &b_{0,i}^2 + \frac{1}{N_k}\sum_{\sigma,k}\langle \chi^{\dagger}_{i,k,\sigma} \chi_{i,k,\sigma}\rangle = 1,\\
        &\frac{1}{N_cN_k} \sum_{\sigma,i,k}\langle c^{\dagger}_{i,k,\sigma} c_{i,k,\sigma}\rangle + \langle \chi^{\dagger}_{i,k,\sigma} \chi_{i,k,\sigma}\rangle = \nu.\\
    \end{split}
\end{equation}
Therefore, the total number of self-consistency equations is $8N_c-2$. 

\begin{figure}
\centering
\includegraphics[width = 0.9\linewidth]{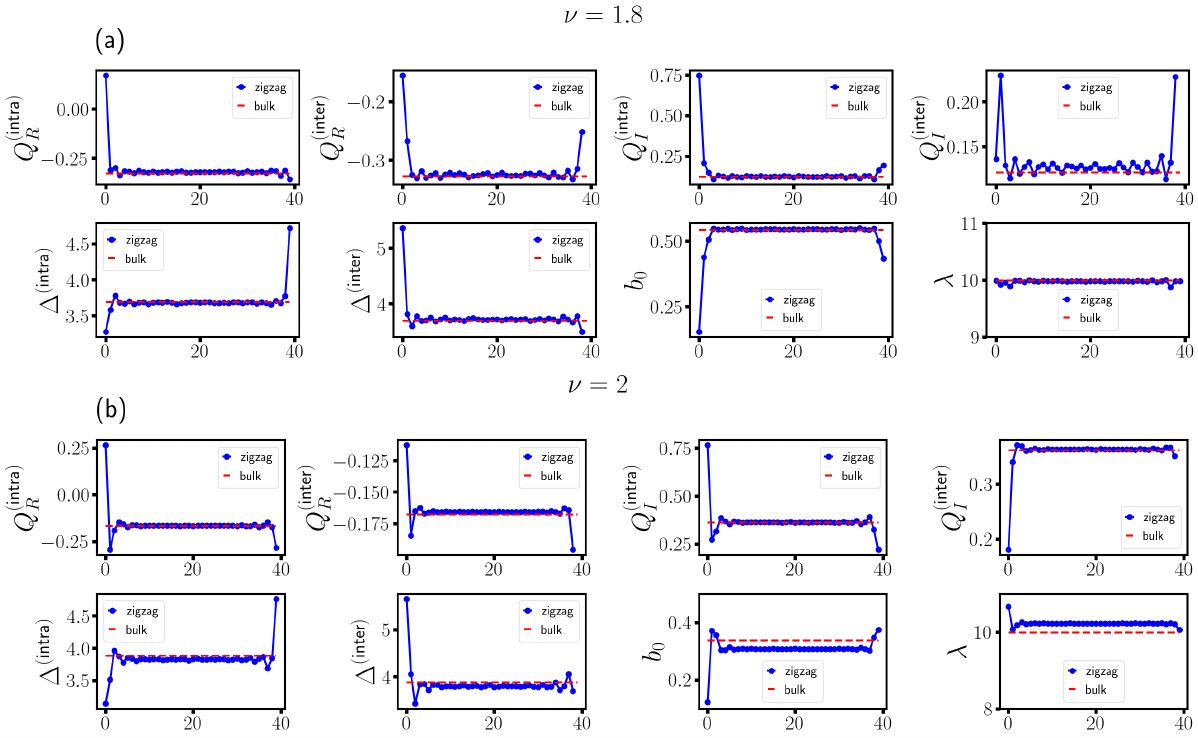}
   \caption{The self-consistency solution in the case of the zigzag boundary condition. The red lines mark the solution in the bulk case. The filling fraction $\nu$ is set to be (a) 1.8 and (b) 2.
   } \label{sol_OBC}
\end{figure}

Fig.~\ref{sol_OBC} displays the self-consistent solution as a function of position. As we can see from these plots, the spatial profile of each variable fluctuates around the bulk solution marked by a red dashed line in each plot. When the position is in the middle region, the solution closely matches the bulk solution. In contrast, the mean field variables exhibit a large deviation near the edges since the finite-size effects are more pronounced. This numerical calculation demonstrates that the solutions still align with the bulk solution in the middle region, and one can expect that the spatial fluctuations will be suppressed as the system size increases, reproducing the mean-field solution in the translationally invariant case. 

Lastly, we explain the numerical procedure of solving the self-consistent equations. The algorithm is based on iteration of the variable. The solution is obtained once the convergence conditions are satisfied. To begin with, we input a set of initial values $X_0$, which is a vector form of variables. Then one can obtain a updated variable$X_{new}$ via self-consistent equations. The convergence is defined as $||X_{new}-X_0||<\epsilon$,where $\epsilon$ is the tolerance taken to be $10^{-4}$. If the above condition is not satisfied, we apply fixed point iteration by mixing the $n-th$ step of the iteration, $X_n$, with the new solution, $X_{new}$, via $X_{n+1}=\alpha X_n+(1-\alpha)X_{new}$ to compute the $n+1$ step of the iteration.

\section{The Landau theory}\label{appD}

\subsection{The free energy expansion}

In this section, we aim to perform the free energy expansion from the microscopic model. We rewrite Eq.~\eqref{free_energy} as a path integral form,

\begin{equation}
    \begin{split}
        Z = \mathcal{N}\int D[c,\bar c,\chi,\bar\chi,b_0,\Delta,Q_\sigma,Q_\sigma^*,\lambda]e^{- S_{0}[c,\bar c,\chi,\bar\chi\psi,b_0,\Delta,Q_\sigma,Q_\sigma^*,\lambda]}, 
    \end{split}
\end{equation}
The action, $S_{0}$, can be obtained from performing a Hubbard-Stratonovich transformation for the interacting fermionic Lagrangian and is expressed as
\begin{equation}
    \begin{split}
        S_0 = \frac{1}{ N_{\bm k}}\sum_{i\omega_n,\bm k,\sigma}\Psi_\sigma^{\dagger}(i\omega_n-H_{\text{MF},\sigma}(\bm k))\Psi_\sigma+ \beta\big( \frac{3|\Delta|^2}{V_1}+ 
        \frac{3\sum_{\sigma}|Q_\sigma|^2}{V_2}+\lambda(1-|b_0|^2)
        \big ).
    \end{split}
\end{equation}
where $\Psi_\sigma = (c_{\bm k,\sigma},\chi_{\bm k,\sigma})^T$ is the spinor for the sublattice degree of freedom for the spin $\sigma$ sector. $\beta$ is inverse temperature. $N_{\bm k}$ denotes the number of momentum mesh points.  To see how does pre-existing SLC emerge as we switch on the inter-layer tunneling, we first derive the effective action by integrating the W degree of freedom. The action reads

\begin{equation}
    \begin{split}
        &\bar S_{eff}[\chi,\bar\chi,b_0,\Delta,Q_\sigma,Q_\sigma^*,\lambda] = 
        \frac{1}{ N_{\bm k}}\sum_{i\omega_n,\sigma}\Psi_\sigma^{\dagger}(G^{-1}_{\chi,\sigma}(i\omega_n,\bm k)-\tilde{\Sigma}_\sigma(\bm k))\Psi_\sigma+ \beta\big( \frac{3|\Delta|^2}{V_1}+ 
        \frac{3\sum_\sigma|Q_\sigma|^2}{V_2}+\lambda(1-|b_0|^2)
        \big ).
    \end{split}
\end{equation}
We define the dressed spinon propagator and self-energies to be
\begin{equation}
    \begin{split}
        G^{-1}_{\chi,\sigma}(i\omega_n,\bm k)&=i\omega_n+\mu-\lambda+\frac{\epsilon_d}{2}-\bar \xi_{\text{M},\bm k}-\Sigma_\sigma(i\omega_n,\bm k),\\
        \Sigma_\sigma(i\omega_n,\bm k) &=  \frac{(-b_0t_\text{h}+\Delta)^2}{i\omega_n-\xi_{\text{W},\bm k,\sigma}+\mu-\frac{\epsilon_d}{2}},\\
        \tilde{\Sigma}_{\sigma}(\bm k) &= Q_\sigma F(\bm k)+Q_{\sigma}^* F^*(\bm k).
    \end{split}
\end{equation}
Here, we have defined $\bar \xi_{\text{M},\bm k}= -2b_0^2t_\text{M}\sum_{i}\cos (\bm k\cdot \bm a_i)$. We further integrate out the spinon degree of freedom, and this operation yields the following effective action

\begin{equation}
    \begin{split}
        S_{eff} = \frac{1}{ N_{\bm k}}\sum_{\bm k,i\omega_n,\sigma}\ln[\det[G^{-1}_{\chi,\sigma}(i\omega_n,\bm k)-\tilde{\Sigma}_\sigma(\bm k)]] +  \beta\big(\frac{3|\Delta|^2}{V_1}+ 
        \frac{3\sum_\sigma|Q_\sigma|^2}{V_2}+ \lambda(1-|b_0|^2)
        \big ).
    \end{split}
\end{equation}
Then the free energy can be obtained via $\mathcal{F}=\frac{S_{eff}}{\beta}$.

To proceed, we expand the logarithm, 
\begin{equation}
    \begin{split}
        &\text{Tr}[\ln[G_{\chi,\sigma}(i\omega_n,\bm k)^{-1} + \tilde{\Sigma}_{\sigma}(\bm k)]]\\
        &= \text{Tr}[\ln[G_{\chi,\sigma}(i\omega_n,\bm k)^{-1}]] + 
        \text{Tr}[\ln[1 + G_{\chi,\sigma}(i\omega_n,\bm k)\tilde{\Sigma}_{\sigma}(\bm k)]]\\
        &= \text{Tr}[\ln[G_{\chi,\sigma}(i\omega_n,\bm k)^{-1}]] + \sum_{m=1}\frac{(-1)^{m-1}}{m}\text{Tr}
        \big[\big(
        G_{\chi,\sigma}(i\omega_n,\bm k)\tilde{\Sigma}_{\sigma}(\bm k)\big)^m\big].\\
    \end{split}
\end{equation}
The first order term for the spin sector $\sigma$ reads

\begin{equation}\label{eq:first}
    \sum_{i\omega_n,\bm k}\text{Tr}[G_{\chi,\sigma}\tilde{\Sigma}_{\sigma}] = 
    \sum_{i\omega_n,\bm k}
    \frac{Q_{\sigma}F(\bm k)+Q_\sigma^*F^*(\bm k)}
    {i\omega_n+\mu-\lambda-\frac{\epsilon_d}{2}-\bar \xi_{\text{M},\bm k,\sigma}-\Sigma_\sigma(i\omega_n,\bm k)}.
\end{equation}
We then substitute $Q_\sigma$ to be $|Q|e^{is_\sigma\phi}$, the above equation becomes
\begin{equation} \label{eq:GL_linear}
\begin{split}
        \sum_{i\omega_n,\bm k}\text{Tr}[G_{\chi,\sigma}\tilde{\Sigma}_{\sigma}] &= 
    \sum_{i\omega_n,\bm k}\frac{1}{i\omega_n+\mu-\lambda-\frac{\epsilon_d}{2}-\bar \xi_{\text{M},\bm k}-\Sigma_\sigma(i\omega_n,\bm k)}|Q|(\sum_i\cos(\bm k \cdot \bm a_i+s_\sigma\phi))\\
    &= 
    \sum_{i\omega_n,\bm k}\frac{1}{i\omega_n+\mu-\lambda-\frac{\epsilon_d}{2}-\bar \xi_{\text{M},\bm k}-\Sigma_\sigma(i\omega_n,\bm k)}|Q| (\sum_i\cos(\bm k \cdot \bm a_i)\cos(s_\sigma\phi)-\sum_i\sin(\bm k \cdot \bm a_i)\sin(s_\sigma\phi))\\
    &= |Q|(\kappa_{1,\sigma}\cos(\phi)-\kappa_{2,\sigma}\sin(\phi)) \\
    & = \kappa_\sigma|Q|\cos(\phi-\theta_\sigma),
\end{split}
\end{equation}
where $\kappa_{1,\sigma},\kappa_{2,\sigma}$ are given by
\begin{equation}\label{eq:kappa}
    \begin{split}
        \kappa_{1,\sigma} = \frac{1}{\beta N_{\bm k}}\sum_{i\omega_n,\bm k}\frac{F_{c}(\bm k)}{i\omega_n+\mu-\lambda-\frac{\epsilon_d}{2}-\bar \xi_{\text{M},\bm k}-\Sigma_\sigma(i\omega_n,\bm k)},\\
        \kappa_{2,\sigma} = \frac{s_\sigma}{\beta N_{\bm k}}\sum_{i\omega_n,\bm k}\frac{F_{s}(\bm k)}{i\omega_n+\mu-\lambda-\frac{\epsilon_d}{2}-\bar \xi_{\text{M},\bm k}-\Sigma_\sigma(i\omega_n,\bm k)}.
    \end{split}
\end{equation}
Here we have defined $\kappa_\sigma = \sqrt{\kappa_{1,\sigma}^2+\kappa_{2,\sigma}^2}$ and $\theta_{\sigma} = -\tan^{-1}(\frac{\kappa_{2,\sigma}}{\kappa_{1,\sigma}})$.  Notice that since the system preserved time-reversal symmetry, one can show that $\kappa_{1,\uparrow} = \kappa_{1,\downarrow}=\kappa_1$ and $\kappa_{2,\uparrow} = \kappa_{2,\downarrow}=\kappa_2$. Similarly, we can also drop the indices on $\kappa_\sigma$ and $\theta_\sigma$. One can find that the expression, Eq.~\eqref{eq:GL_linear}, is consistent with the first term in Eq.~\eqref{eq:GLFE} in the main text. Subsequently, we also derive the expression for the coefficient $\alpha$ for the quadratic term

\begin{equation}
    \begin{split}
        \alpha &=  
        -\frac{2}{\beta N_{\bm k}}\sum_{i\omega_n,\bm{k}} \big((\frac{1}{i\omega_n-\mu+\frac{\epsilon_d}{2}-\bar\xi_{\text{M},\bm k}-\Sigma_\sigma(i\omega_n,\bm k)})^2
(F_c(\bm{k})^2+F_s(\bm{k})^2)\big )+\frac{3}{V_2}.
    \end{split}
\end{equation}
Again, the time reversal symmetry suggests that the value of $\alpha$ is identical in both spin-up and spin-down sectors.  
\begin{figure}[h]
\centering
   \includegraphics[width=0.8\linewidth]{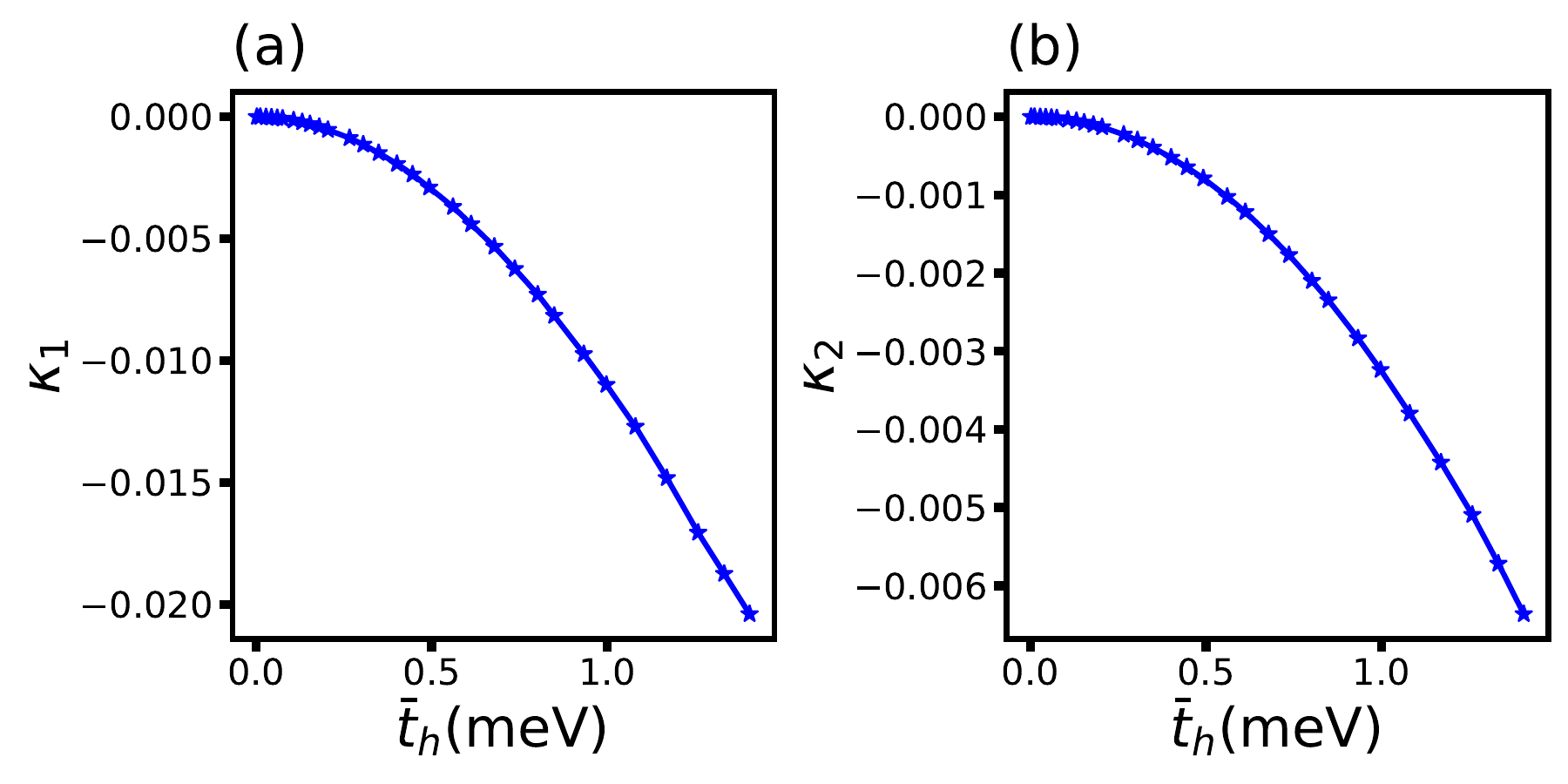}
   \caption{The real part ($\kappa_1$) and imaginary part ($\kappa_2$) of the coefficient $\kappa$ as a function of total hybridization $\bar{t}_{h}$ in (a) and (b), respectively. The parameters for the self consistency are chosen to be: $ t_{\text{W}}=9\text{ meV},\;t_{\text{M}}=4.5\text{ meV},\;\epsilon_d=3\text{ meV}$, and $ V_2=0\text{ meV},T = 0.01\text{ meV} $. Computed values are shown by star markers with line drawn as a guide for the eye.}\label{coefficients}
\end{figure}

In general, both $\kappa_1$ and $\kappa_2$ are finite. Only in the decoupled limit where inter-layer tunneling is absent does $\kappa_2$ vanish due to symmetry because the form of the function is odd. In addition, a finite inter-layer tunneling renders the  spinon propagator to be neither even or odd in momentum since the self-energy $\tilde{\Sigma}(\bm k) \neq \pm\tilde{\Sigma}(-\bm k)$.

Now we continue our calculation by re-arranging Eq.~\eqref{eq:kappa} as follows
\begin{equation}
    \begin{split}
        \kappa_{1/2} = \frac{1}{N_{\bm k}\beta}\sum_{i\omega_n,\bm k}
        \Big[\big(\frac{1}{2}-\frac{\xi_{W,\bm k,\sigma}-\bar\xi_{M,\bm k}-\lambda+\epsilon_d}{E_{+,\bm k,\sigma}-E_{-,\bm k,\sigma}}\big)F_{c/s}(\bm k)\frac{1}{i\omega_n-E_{+,\bm k,\sigma}}+ \\
        \big(\frac{1}{2}+\frac{\xi_{W,\bm k,\sigma}-\bar\xi_{M,\bm k}-\lambda+\epsilon_d}{E_{+,\bm k,\sigma}-E_{-,\bm k,\sigma}}\big)F_{c/s}(\bm k)\frac{1}{i\omega_n-E_{-,\bm k,\sigma}}\Big],
    \end{split}
\end{equation}
where $E_{\pm,\bm k,\sigma}$ is defined as 
\begin{equation}
    \begin{split}
        E_{\pm,\bm k,\sigma} = \frac{\xi_{\text{W},\bm k,\sigma}+\bar\xi_{\text{M},\bm k}-2\mu+\lambda}{2}\pm
        \sqrt{(\frac{\xi_{\text{W},\bm k,\sigma}-\bar\xi_{\text{M},\bm k}+\epsilon_d-\lambda}{2})^2+(b_0t_\text{h}-\Delta)^2|V(\bm k)|^2}.
    \end{split}
\end{equation}
The frequency sum for $\kappa_{1/2}$ yields:
\begin{equation}\label{eq:linear_GL_coeff}
    \begin{split}
        \kappa_{1/2} = \frac{1}{N_{\bm k}}\sum_{\bm k}\Big[
        \big(\frac{1}{2}-\frac{\xi_{W,\bm k,\sigma}-\bar\xi_{M,\bm k}-\lambda+\epsilon_d}{E_{+,\bm k,\sigma}-E_{-,\bm k,\sigma}}\big)F_{c/s}(\bm k)f(E_{+,\bm k,\sigma})+
        \big(\frac{1}{2}+\frac{\xi_{W,\bm k,\sigma}-\bar\xi_{M,\bm k}-\lambda+\epsilon_d}{E_{+,\bm k,\sigma}-E_{-,\bm k,\sigma}}\big)F_{c/s}(\bm k)f(E_{-,\bm k,\sigma})
        \Big].
    \end{split}
\end{equation}
Here, $f(E)$ is the Fermi function. Following in a similar manner, the expression for the quadratic coefficient, $\alpha$, can be evaluated:

\begin{equation}
    \begin{split}
        \alpha = -\frac{\beta}{2N_{\bm k}}\sum_{\bm k}\cosh^{-1}\big(\frac{\beta(\mu-\frac{\epsilon_d }{2}-\bar\xi_{\text{M},\bm k})}{2}\big)(F_c(\bm{k})^2+F_s(\bm{k})^2)+\frac{3}{V_2}.
    \end{split}
\end{equation}

To demonstrate the presence of $\kappa_1$ and $\kappa_2$,  we numerically evaluate both as functions of total inter-layer tunneling $\bar t_\text{h}= \Delta-b_0 t_{\text{h}}$ which are shown in Fig.~\ref{coefficients}. In particular, we obtain the mean field parameters self-consistency and put their solutions into the evaluation of the coefficients in Eq.~\eqref{eq:linear_GL_coeff}. To vary $\bar t_\text{h}$, we solve the self-consistency from $V_1=0\text{ meV}$ to $V_1=5\text{ meV}$. We also set $V_2$ to zero to enforce the fact that the SLC is intrinsic to the model and generated by inter-layer tunneling and spin-orbit coupling on the W triangular lattice rather than an effect of the interactions alone. From the plots, we found that, both $\kappa_1$ and $\kappa_2$ become vanishingly small as $\bar t_\text{h} \ll 1$. For $\kappa_1$, its value is determined by the effective hopping strength between Mo sites. When $\bar t_{\text{h}}$ is small, the boson condensation $b_0$ is also small, rendering the hopping between Mo sites $b_0^2t_{\text{M}}$ to be suppressed. Consequently, $\kappa_1$ is also suppressed due to small boson condensation.  In addition, one can also observe a similar behavior in $\kappa_2$. As the inter-layer tunneling gets suppressed, there is little pre-existing SLC on the Mo layer induced by the W layer and thereby lower the value of $\kappa_2$. In the decouple limit, $\kappa_2$ would vanish, consistent with the symmetry argument made just prior.

We lastly note that the above Ginzburg-Landau theory is formulated within a uniform gauge, raising an issue of what happens if the gauge is not fixed. In general, the phase of $Q$ is not physically observable given its gauge-dependent behavior (see Eq.~\eqref{eq:gauge_bond_fields}). However, the induced SLC on the Mo layer via inter-layer hybridization will generate a bond current (following the definition in the main text) that remains unaffected by the choice of gauge. The emergence of this current leads to a complex bond field $Q$, whose phase structure is likely constrained by time-reversal symmetry and the space group symmetry of the moir\'{e} lattice. This consideration motivates the ansatz taken in our uniform gauge. We emphasize that the phase of the field $Q$ is not gauge invariant, but it's contribution to the bond current, as defined in Ref.~\cite{gleis2026topologicalkondoinsulatorspin}, is invariant under a gauge transformation.

\section{Analytical treatment of zigzag edge state velocity}\label{appE}

\begin{figure}\label{ladder}
\centering
   \includegraphics[width=0.8\linewidth]{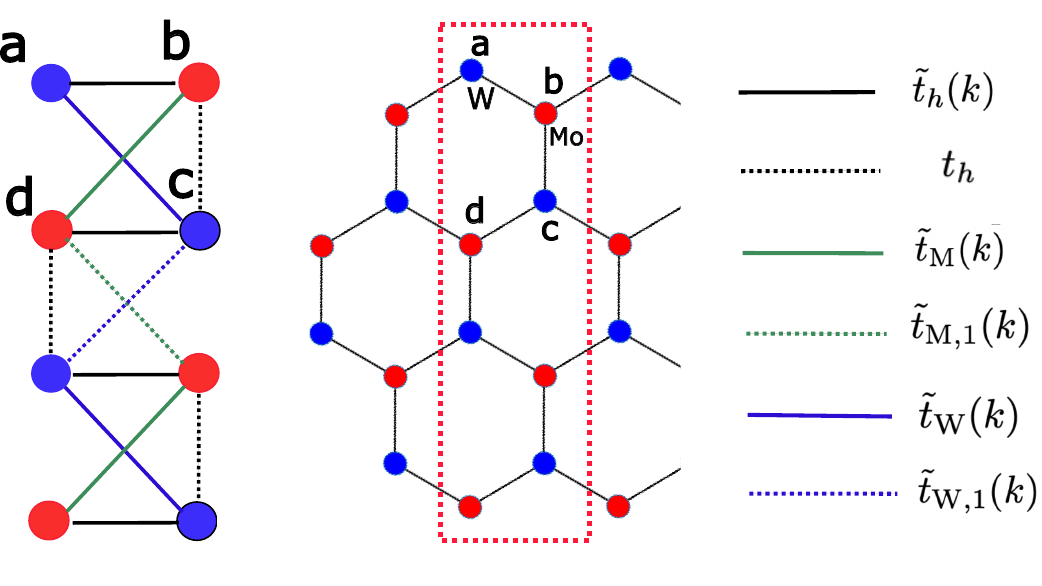}
   \caption{Schematic illustration of the simplified geometry and momentum-dependent hopping processes. The left panel shows the effective ladder representation of the region enclosed by the red dashed rectangle in the honeycomb lattice shown in the middle panel. Blue and red circles denote W and Mo sites, respectively, while $a$, $b$, $c$, and $d$ label the four sites within the unit cell. Different line colors and styles distinguish the hopping processes, as indicated in the legend on the right.}\label{ladder}
\end{figure}

\begin{figure}[h]
\centering
   \includegraphics[width=0.7\linewidth]{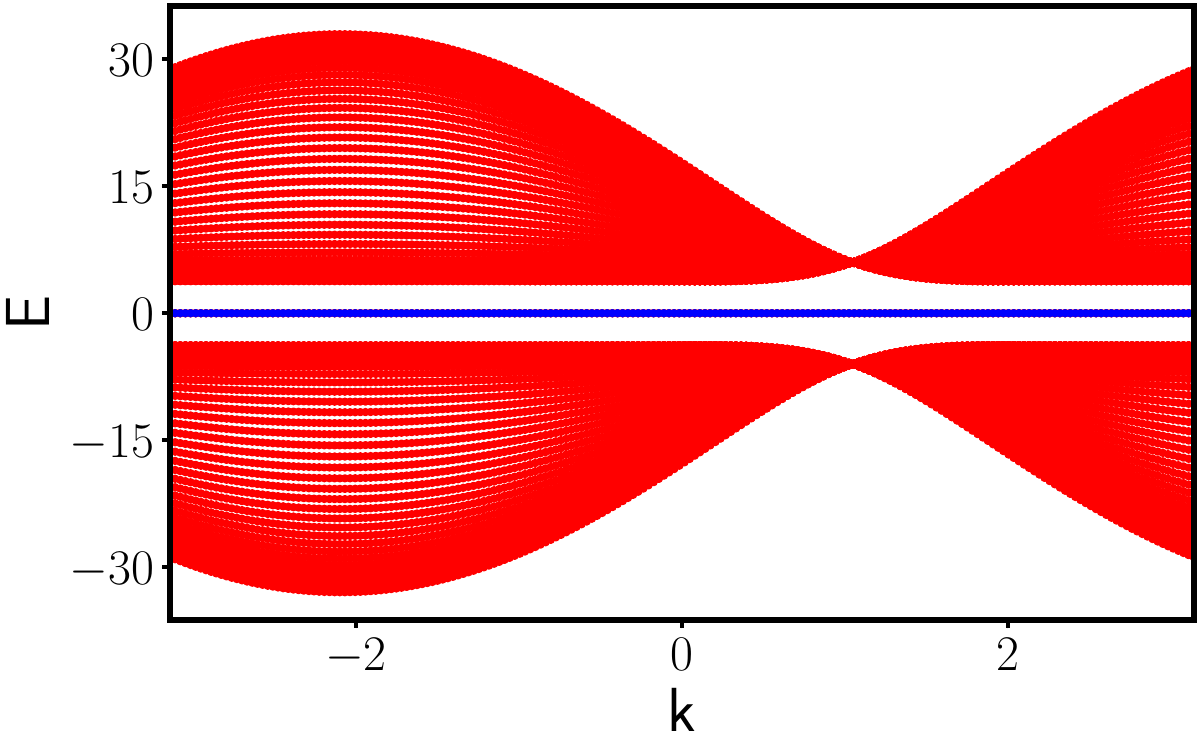}
   \caption{Representative band structure of $H_0$. The parameters are chosen to be $t_\text{W}=8,t_\text{M}=1,t_\text{h}=6,\phi_\text{W} = \frac{2\pi}{3},\phi_\text{M} = -\frac{\pi}{3}$.
   From the band structure, one can see the flat-band through the whole Brillioun zone.
   }\label{flat_spectrum}
\end{figure}

In this Appendix, we present the analytical treatment of the edge-state velocity on the zigzag boundary condition based on Ref.~\cite{daiTwodimensionalWeylNodalline2024,rahmatiExplicitDerivationChiral2023}. The central idea is to separate the Hamiltonian into an unperturbed part and a perturbed part. For this purpose, one needs to redefine the unit-cell with four atoms, labeled by $a,b,c$ and $d$ (see Fig.~\ref{ladder}). Inheriting the notation in Appendix~\ref{appB}, we can rewrite the model Hamiltonian based on the enlarged unit-cell
\begin{equation}
    \begin{split}
    \hat{H} = \hat{H}_0+\hat{H}_{\text{pert}},\\
    \end{split}
\end{equation}
with 
\begin{equation}
    \begin{split}
          \hat{H}_0 = \sum_{i}\Psi_{k,i}^\dagger T_0(k)\Psi_{k,i} + \Psi_{k,i+1}^\dagger T_1(k)\Psi_{k,i}.
    \end{split}
\end{equation}
and
\begin{equation}
    \hat{H}_{\text{pert}} =  \sum_{i}\Psi_{k,i}^\dagger T_2(k)\Psi_{k,i}.
\end{equation}
$T_0,T_1,T_2$ are defined as 

\begin{equation}
    \begin{split}
        T_0(k) &= 
        \left(
        \begin{array}{cccc}
             0&  0& \tilde{t}_{\text{W}}(-k)&0\\
            0 &  0& t_h& \tilde{t}_{\text{M}}(k)\\
            \tilde{t}_{\text{W}}(-k)&t_h&0&0\\
            0&\tilde{t}_{\text{M}}(k)&0&0
        \end{array}
        \right),\\
         T_1(k) &= 
          \left(
        \begin{array}{cccc}
             0&  0& \tilde{t}_{\text{W}}(-k)&t_h\\
            0 &  0& 0& \tilde{t}_{\text{M}}(k)\\
            0&0&0&0\\
            0&0&0&0
        \end{array}
        \right),\\
        T_2(k) &= 
          \left(
        \begin{array}{cccc}
             \tilde{t}_{\text{W},1}(k) &  \tilde{t}_{h}(k)& 0&0\\
             \tilde{t}_{h}(k) &   \tilde{t}_{\text{M},1}(-k)& 0& 0\\
            0&0&\tilde{t}_{\text{W},1}(k) &  \tilde{t}_{h}(k)\\
            0&0&  \tilde{t}_{h}(k) &   \tilde{t}_{\text{M},1}(-k)
        \end{array}
        \right),\\
    \end{split}
\end{equation}
with

\begin{equation}
\begin{split}
    \tilde{t}_{\text{W}}(k) &= 2t_\text{W}\cos (\frac{k}{2}-\phi_\text{W}),\\
    \tilde{t}_{\text{M}}(k) &= 2t_\text{M}\cos (\frac{k}{2} -\phi_\text{M}), \\
    \tilde{t}_{\text{W},1}(k) &= 2t_\text{W}\cos (k-\phi_\text{W}),\\ 
    \tilde{t}_{\text{M},1}(k) &= 2t_\text{M}\cos (k-\phi_\text{M}), \\
    \tilde{t}_h(k) &= 2t_h\cos(\frac{k}{2}).  
\end{split}
\end{equation}
Here, $\phi_\text{W}, \phi_\text{M}$ denote the phase of the hopping processes. The Hamiltonian matrix is written in the basis $(\psi_{a,i,k})^T$ and hopping processes $\tilde{t}_{\text{W}}(k),\tilde{t}_{\text{W},1}(k),\tilde{t}_{\text{M}}(k),\tilde{t}_{\text{M},1}(k),\tilde{t}_{h}(k)$ and $t_h$ are graphically presented in the left-most cartoon in Fig.~\ref{ladder}. In the following derivation, for convenience, we will drop the k dependence in notation. The analytical treatment relies on numerical observation. We first numerically diagonalized the unperturbed Hamiltonian $\hat{H}_0$ and found a pair of zero modes for any momentum in the Brillouin zone, as shown in Fig.~\ref{flat_spectrum}.

Here, we note that the two-fold degenerate zero modes do not persist upon breaking the chiral symmetry. For example, if we apply the staggered potential, which corresponds to $\lambda$ and the displacement field $\epsilon_d$ in a realistic setup, it can lift the degeneracy and make the flat band dispersive. As a consequence, this methodology is invalid once the staggered potential is prominent. However, if we consider the small displacement field limit, this analytical treatment is still applicable.

To proceed, let us gain insight into the eigenfunction of the zero modes.  The two localized edge states only have two non-vanishing components in the wavefunction, as $H_0$ has a Chiral symmetry.  Therefore, we assume a wavefunction ansatz in which the site $c$ and $d$ has zero amplitudes. The edge state wavefunction can be written as
\begin{equation}
    |\Psi_{\text{edge}}\rangle = \bigoplus_i (\psi_{a,i},\psi_{b,i},0,0)^T.
\end{equation}
With the given ansatz and zero energy eigenvalues, the eigenvalue equation is given by

\begin{equation}
    \begin{split}
        \left(
        \begin{array}{cc}
            \tilde{t}_\text{W} & t_\text{h} \\
            0 & \tilde{t}_\text{M}
        \end{array}
        \right)
         \left(
        \begin{array}{c}
            \psi_{a,i} \\
             \psi_{b,i}
        \end{array}
        \right)
        +
        \left(
        \begin{array}{cc}
            \tilde{t}_\text{W} & 0 \\
            t_\text{h} & \tilde{t}_\text{M}
        \end{array}
        \right)
         \left(
        \begin{array}{c}
            \psi_{a,i+1} \\
             \psi_{b,i+1}
        \end{array}
        \right)
        =0.
    \end{split}
\end{equation}
The above equation relates the wavefunction of site index $i$ and that of index $i+1$. This recursion relation yields the following equation

\begin{equation}
    \begin{split}
        \begin{pmatrix}
             \psi_{a,n+1}  \\
             \psi_{b,n+1}
        \end{pmatrix}
        =M^n
        \begin{pmatrix}
             \psi_{a,1}  \\
             \psi_{b,1}
        \end{pmatrix},
    \end{split}
\end{equation}
where the matrix $M$ is defined as

\begin{equation}
    \begin{split}
        M &= -T_1^{-1}T_0
        =
         \left(
        \begin{array}{cc}
            -1 & -\frac{\tilde{t}_\text{h} }{\tilde{t}_\text{W}} \\
            \frac{\tilde{t}_\text{h} }{\tilde{t}_\text{M}} &
            -1+\frac{\tilde{t}_\text{h} ^2}{\tilde{t}_\text{W}\tilde{t}_\text{M}}
        \end{array}
        \right).
    \end{split}
\end{equation}
The eigenvalues of the matrix $M$ can be easily computed analytically and are given by

\begin{equation}
    \begin{split}
        \lambda_{\pm} = -1+\frac{\tilde{t}_\text{h}^2}{2\tilde{t}_\text{W}\tilde{t}_\text{M}}
        \pm\frac{\tilde{t}_\text{h}^2}{\tilde{t}_\text{W}\tilde{t}_\text{M}}\text{sgn}(\tilde{t}_\text{W}\tilde{t}_\text{M})\sqrt{\frac{1}{4}+\frac{\tilde{t}_\text{W}\tilde{t}_\text{M}}{\tilde{t}_\text{h}^2}}.
    \end{split}
\end{equation}
To obtain a localized solution, we focus on cases with $|\lambda_{\pm}|<1$. We then write down the eigenstates:

\begin{equation}\label{eq:edge_wavefn}
    |\Psi_{\text{edge},L}\rangle = 
    \frac{1}{C_L}\left(
    \begin{array}{c}
       \frac{\tilde{t}_\text{h}}{\tilde{t}_\text{W}}
       \big(\frac{1}{2}-\sqrt{\frac{1}{4}-\frac{\tilde{t}_\text{W}\tilde{t}_\text{M}}{\tilde{t}_\text{h}^2}}\big) \\
       1
    \end{array}
    \right),
    \quad 
    |\Psi_{\text{edge},R}\rangle = 
    \frac{1}{C_R}\left(
    \begin{array}{c}
       1 \\
        \frac{\tilde{t}_\text{h}}{\tilde{t}_\text{M}}
       \big(\frac{1}{2}-\sqrt{\frac{1}{4}-\frac{\tilde{t}_\text{W}\tilde{t}_\text{M}}{\tilde{t}_\text{h}^2}}\big)
    \end{array}
    \right).
\end{equation}
where $C_{L/R}^2 = 1+   \big(\frac{\tilde{t}_\text{h}}{\tilde{t}_\text{W/M}}
       \big(\frac{1}{2}-\sqrt{\frac{1}{4}-\frac{\tilde{t}_\text{W}\tilde{t}_\text{M}}{\tilde{t}_\text{h}^2}}\big)  \big)^2$  are normalization factors. Interestingly, Taking $t_{\text{M}} \rightarrow 0$ in Eq.~\eqref{eq:edge_wavefn} yields the wavefunction for the right edge state
$|\Psi_{\text{edge},L}\rangle = \left(\begin{array}{c}
    1  \\
     0
\end{array}\right)$. One can see that the wavefunction weight only concentrates on Mo sites. A similar behavior can also be seen in $t_\text{W}\rightarrow 0$ case.

       With the closed form of the wavefunctions, we can derive via first-order perturbation theory the edge state energy, $E_a(k)$ as:
       
\begin{equation}
\begin{split}
       E_a(k) = \langle \Psi_{\text{edge},a}|H_{\text{pert}}|\Psi_{\text{edge},a}\rangle,\\
\end{split}
\end{equation}
where $a = L,R$. To obtain a simple relation, we first consider the limit where $t_\text{W},t_\text{M}<t_\text{h}$.  In this case, the normalization factors can be approximated as $C_{L/R}^2\approx1+4(\frac{\tilde{t}_{\text{W/M}}}{\tilde{t}_\text{h}})^2$. The edge state velocities are given by

\begin{equation}\label{eq:edgevelocity}
    \begin{split}
        V_{L/R} = \frac{6t_{\text{M/W}}\sin(k+\phi_{\text{M/W}})}{\sqrt{1+4(\frac{\tilde{t}_{\text{M/W}}}{\tilde{t}_\text{h}})^2}}.
    \end{split}
\end{equation}
It is interesting to discuss the case where $t_{\text{M}}$ is zero. Based on  Eq.~\eqref{eq:edgevelocity}, the velocity at right edge disappears. The vanishing velocity can be understood from the wavefunction of the right edge state, leading to the vanishing velocity along the edge direction.

\section{Derivation of self-energy matrix within the wide band approximation}

\begin{figure}[h]
\centering
   \includegraphics[width=0.5\linewidth]{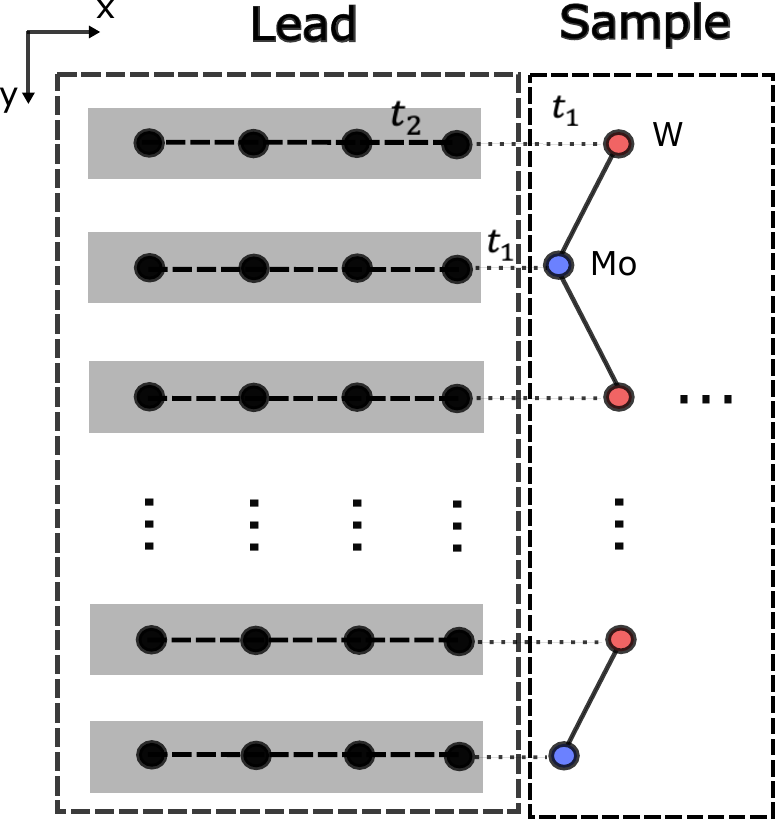}
   \caption{Schematic illustration for the connection between the lead and the TMD sample.}\label{lead}
\end{figure}

In the conductance calculation, we model the self-energy matrix as an imaginary on-site potential on the interface between the TMD sample and leads. Here, I provide a derivation of self-energy matrix based on a simple microscopic model. To begin with, we treat the lead as a composition of multiple one-dimensional chains, each of which couple to one particular site on the edge of the TMD sample, see Fig.~\ref{lead}. Since the coupling is spin independent, I ignore the spin degree of freedom in the later derivation. The full model Hamiltonian is given by 
\begin{equation}
    \begin{split}
        H = H_{TMD} + \sum_{a = L/R}H_{a-TMD} + \sum_{a=L/R}H_a,
    \end{split}
\end{equation}
where $H_{TMD}$ is the Hamiltonian of the sample given by the mean field Hamiltonian Eq.~\eqref{eq:full_Ham}.  $H_a$ is the lead Hamiltonian which, in the momentum space, can be expressed as
\begin{equation}
    \begin{split}
        H_a = \sum_{k,i}2t_2\cos(k)|a,k,i\rangle\langle a,k,i|,
    \end{split}
\end{equation}
where $a = L/R$ denotes left or right terminal. The coupling between the lead and the sample reads
\begin{equation}
    \begin{split}
        H_{a-TMD} = \sum_{k}t_1|a,k,i\rangle\langle \text{Mo},a,i|+\sum_{k}t_1|a,k,i\rangle\langle \text{W},a,i|.
    \end{split}
\end{equation}
Here, I use basis $|\text{W}(\text{Mo}),L/R,i\rangle$ represents the single particle state on the left/ right edge of the sample with W(Mo) site index $i$. After integrating out the lead degree of freedom, one can obtain the following effective Hamiltonian,
\begin{equation}
    \begin{split}
        H^{(eff)}_{TMD} = H_{TMD} + \Sigma(\omega).
    \end{split}
\end{equation}
With $\Sigma(\omega)$ being the self-energy matrix. Furthermore, the self-energy can be decomposed into a right part $\Sigma_R(\omega)$ and a left part $\Sigma_L(\omega)$. Each of them can be expressed as

\begin{equation}
    \begin{split}
        \Sigma_a(\omega) &= H_{a-TMD}\frac{1}{\omega-H_a}H^\dagger_{a-TMD}\\
        &= t_1^2\sum_{\alpha ,k,i}\frac{1}{\omega-2t_2\cos(k)}|\alpha ,a,i \rangle\langle \alpha ,a,i|\\
        &=\tilde{\Sigma_a}(\omega)\sum_{\alpha ,i}|\alpha ,a,i \rangle\langle \alpha ,a,i|,
    \end{split}
\end{equation}
where $\tilde{\Sigma_a}(\omega) = t_1^2\rho(0)\int^{D}_{-D}d\epsilon\frac{1}{\omega-\epsilon} $ is a complex number. Here, we have converted the momentum sum to an energy integral, with a factor $\rho(0)$ being the density of state at the Fermi level. To proceed, Notice that 
\begin{equation}
    \begin{split}
        \frac{1}{\omega-\epsilon} = \mathcal{P}(\frac{1}{\omega-\epsilon})-i\pi\delta(\omega-\epsilon),
    \end{split}
\end{equation}
where $\mathcal{P}(...)$ denotes Cauchy principal value. 

\begin{figure}[h]
\centering
   \includegraphics[width=0.6\linewidth]{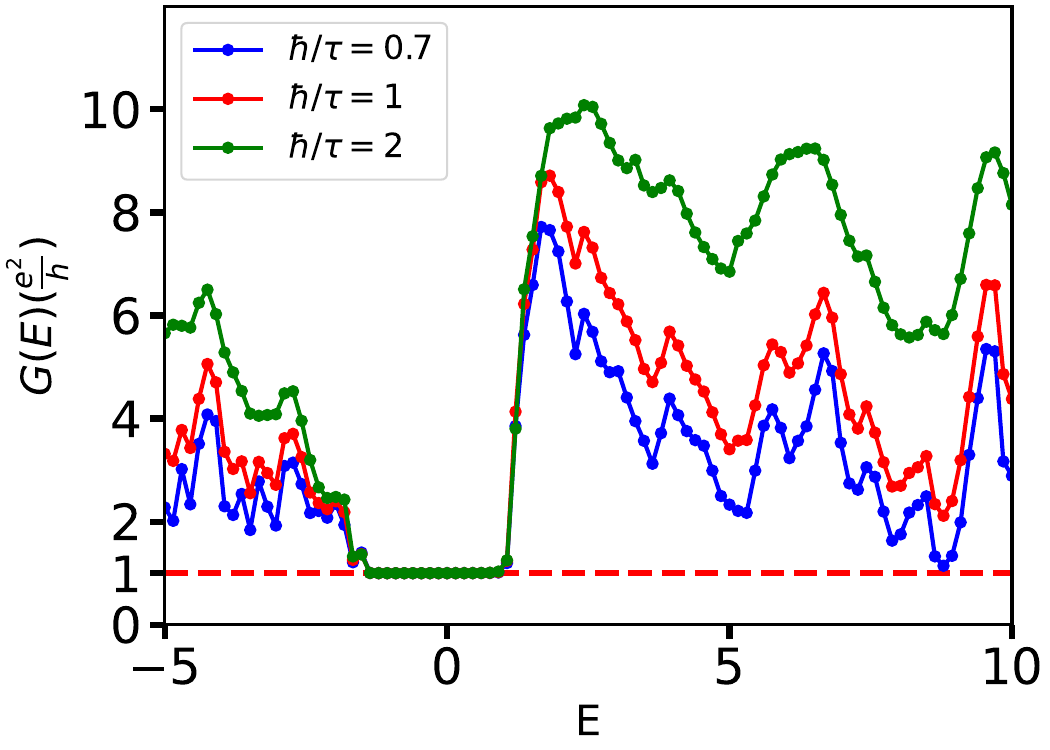}
   \caption{Conductance as a function of energy. Three different coupling strengths ($\hbar/\tau = 0.7,1,2 \text{ meV}$) are considered. The model parameters used in the computation is the same as Fig.~\ref{conductance}~(d). }\label{conductance_couple}
\end{figure}

For the imaginary part of  $\Sigma(\omega)$ , the energy integration yields $\text{Im}[\Sigma(\omega)] = -\pi t_1^2\rho(0)$. For the real part of $\Sigma(\omega)$ , we have 
\begin{equation}
    \begin{split}
        \text{Re}[\Sigma(\omega)] = t_1^2\rho(0)\ln\big|\frac{\omega-D}{\omega+D}\big|.
    \end{split}
\end{equation}
In the wide band limit where $D\gg \omega$ , $\text{Re}[\Sigma(\omega)]$ goes zero. Thus, the self energy is purely imaginary and energy independent. We thus have $\Sigma(\omega) = -i\pi t_1^2\rho(0)\sum_{a=W,Mo}|a,i\rangle\langle a,i|$. Connecting to Eq.~\eqref{eq:sigma}  in the main text, we have $\hbar/\tau = \pi t_1^2\rho(0)$ and treat it as a phenomenological parameter. Correspondingly, the effective coupling $\Gamma_{L/R}$ is proportional to $\hbar/\tau$. 

Lastly, let us discuss the effect of coupling strength $\Gamma_{L/R}$. Fig.~\ref{conductance_couple} presents three  curves with different coupling strengths. One can see that as the coupling goes up, the conductance of the in-gap region remains to be $\frac{e^2}{h}$, indicating that the conductance from edge transport is independent of the coupling strengths. While outside the in-gap region, the conductance goes up as the coupling strength gets larger. This behavior can be understood from the Landauer-Buttiker formula Eq.~\eqref{Conductance}. As the coupling becomes stronger, the transmission also gets larger, leading to a larger conductance.

\twocolumngrid
\bibliographystyle{apsrev4-2}
\bibliography{TMD}

\end{document}